\documentclass[12pt]{article}
\usepackage{amscd, amssymb}
\usepackage[mathscr]{eucal}
\newtheorem{thm}{Theorem}
\newtheorem{lem}{Lemma}
\newtheorem{ex}{Example}
\newtheorem{defn}{Definition}

\newtheorem{rem}{Remark}

\newcommand{\eep}{\hfill $\square$}
\newcommand{\pf}{\noindent {\bf Proof. \ }}
\usepackage{epsfig}
\title{\Large Quasi Cyclic LDPC Codes Based on Finite Set
Systems}
\author{Mohammad Gholami\thanks{Corresponding author: Mohammad Gholami is with the Department of Mathematics,
Shahrekord University, 115, Shahrekord, IRAN, e-mail: gholami-m@sci.sku.ac.ir, gholamimoh@gmail.com.},~Mehdi Samadieh\thanks{Researcher of Isfahan Mathematics House, Isfahan, IRAN, e-mail:m.samaieh@mathhouse.org samadieh.m@gmail.com.}, 
 }
= 6.7in
\begin{document}
\maketitle
\begin{abstract}
A finite set system (FSS) is a pair $(V,{\cal B})$ where $V$ is a
finite set whose members are called points, equipped with a finite
collection of its subsets $\cal B$ whose members are called
blocks. In this paper, finite set systems are used to define a
class of Quasi-cyclic low-density parity-check (LDPC) codes,
called FSS codes, such that the constructed codes possess large
girth and arbitrary column-weight distributions. Especially, the
constructed column weight-2 FSS codes have higher rates than the
column weight-2 geometric and cylinder-type codes with the same
girths. To find the maximum girth of FSS codes based on $(V,{\cal
B})$,  {\it inevitable walks} are defined in ${\cal B}$ such that
the maximum girth is determined by the smallest length of the
inevitable walks in $\cal B$. Simulation results show that the
constructed FSS codes have very good performance over the AWGN
channel with iterative decoding and achieve significantly large
coding gains compared to the
random-like LDPC codes of the same lengths and rates.
\end{abstract}
\section{Introduction}
Low-density parity-check (LDPC) codes are forward error-correction
codes, first proposed in 1962 by Gallager~\cite{gal} and rediscovered~\cite{mac} in 1996.
The construction methods of LDPC codes can be divided into two
categories: random-like methods~\cite{rand1}-\cite{rand5} and mathematically structured
methods~\cite{struct1}-\cite{protograph}. Long random-like LDPC codes in
general perform closer to the Shannon limit~\cite{shan} than their equivalent
structured LDPC codes; however, for practical lengths, well
designed structured LDPC codes show better error correcting
performance than the random-like ones. Quasi-cyclic (QC) LDPC codes are
the most promising class of structured LDPC codes due to their
ease of implementation and excellent performance over noisy
channels when decoded by message-passing algorithms, such as sum-product algorithm~\cite{sum}, as extensive
simulation studies have shown. An LDPC code is represented by a
sparse parity-check matrix and its corresponding Tanner graph~\cite{tan1}.
Tanner gave a lower bound on the minimum
distance of a given LDPC code that grows exponentially with the girth of the Tanner graph
representing the code.
\medskip

PEG algorithm~\cite{peg} is a graph conditioning technique to construct Tanner graphs of LDPC codes which possess large girths. The algorithm builds a Tanner graph by connecting the graphs nodes edge by edge provided that the added edge has minimal impact on the girth of the graph. Although PEG codes are among the best codes at short lengths~\cite{peg}, their major disadvantage is represented by their high implementation complexity that makes them impractical at very large lengths. On the other hand, structured LDPC codes, especially quasi-cyclic (QC) LDPC codes, such as FSS codes, have advantages over other types of LDPC codes in hardware implementation of encoding and decoding.
\medskip

The overall bit-error rate (BER) and block-error rate (BLER) performance of an LDPC
code is generally described by two different regions, i.e the
waterfall region (WR) and the error-floor region (EFR). The WR
corresponds to the low-to-medium signal-to-noise ratio (SNR) of the BER-SNR plot, and
EFR~\cite{ber} is located at the bottom of WR wherein the BER/BLER no longer exhibits the rapid
improvement as in WR. While the performance in EFR mainly depends on
the minimum-distance, stopping sets~\cite{stoping} and
trapping sets~\cite{traping}, the girth influences the achievable
BER/BLER in WR. On the other hand, the higher the girth, the faster the
iteration-aided BER/BLER improvement, and this is why many
construction techniques attempt to maximize the girth of the
underlying graph~\cite{bone}.
Also, it has been shown that designing column-weight two
LDPC codes~\cite{ghol2,ghol3} with large girths, especially in the non-binary setting, is
highly beneficial for the error floor performance.
\medskip

Moura~\cite{moura} defined  a {\it structure graph} for column-weight
two LDPC codes, or {\it cycle codes}~\cite{cyclecodes},
to construct regular structured LDPC codes with large girth.
Asamov and Aydin~\cite{asamov} presented a greedy algorithm to
construct regular cycle codes with arbitrarily girth. In
\cite{sun}, the authors have viewed a QC-LDPC code as a protograph
code with circulant permutation matrices. They have also proposed
a new combinatorial method for the construction of protographs
whose protograph codes have girth larger than or equal to 14
and 18 for non applicable lengths. Also, a class of geometrically structured QC-LDPC codes with
girth at most 18~\cite{ghol4} is constructed based on some
algebraic tools, i.e. Steiner triple systems, integer lattices, and
affine planes. Furthermore, a particular class of circulant LDPC codes
referred to as cylinder-type LDPC (CT-LDPC) codes has been studied in~\cite{ghol2} wherein for each $e\ge 2$,
column-weight two CT-LDPC codes with girth $8e$ and rate $1/e$ have been constructed.

\medskip

Despite having a suitable theoretical perspective for the mentioned constructions,
it is desired to have efficient deterministic algorithms to produce  LDPC codes of acceptable length, with
flexible parameters such as rate, girth and row (column) weight distribution.
In this paper, by introducing and applying a concept referred to as {\it inevitable walk},
we use the smallest length of such inevitable walks to give an efficient algorithm finding an upper-bound on the girth of QC-LDPC codes having a fixed mother matrix. Then, we introduce an efficient algorithm to produce mother matrices of large girths whose parity-check matrices have desirable column-weight distributions. As an advantage of this algorithm, the so constructed column-weight two LDPC codes have rates larger than the rates of the column-weight two geometric and cylinder-type codes with the same girths given in \cite{ghol2} and \cite{cylinder}. The basic mathematical structure employed in this paper is the class of finite set systems.
\medskip

The outline of the paper is as follows. In section 2, we give some preliminaries and notations.
To a circulant code, a graph, called {\it block-structure graph} (BSG)~\cite{cylinder}, is associated which is simpler than the graphs such as
protograph~\cite{protograph} and voltage-graph~\cite{bocharova}.
By assigning a new matrix $H$ to a given FSS, we define its corresponding FSS-codes as the QC-LDPC codes with mother matrix $H$.
Construction of FSS-based LDPC codes is addressed in Section 4.
By defining {\it inevitable walks} in a FSS, and by using BSG, we give a theorem which determines the maximum achievable girth by FSS-based
codes derived from a given mother matrix. In section 4, we introduce two efficient algorithms to produce regular and non-regular FSS-based codes with desirable girths and column-weight distributions. 
Then, we propose two efficient algorithms producing regular FSS-based codes with arbitrary girths and column-weight distributions.
Simulation results, given in Section 5, show that the
new FSS-based codes can perform very well over the
AWGN channel with iterative decoding, and they achieve a better coding gain compared to
the random codes of the same lengths and rates.
\section{Preliminaries and Notations}
A {\it graph} $G$ is a two-tuple consisting of a {\it vertex set}
$V(G)$ and an {\it edge set} $E(G)$, where each element of $e\in E(G)$ is represented by
two elements, not necessarily distinct,
of $V(G)$ called the endpoints of this edge. A {\it bipartite graph} is a graph $G=(V,E)$ in which $V$ can be
divided into two disjoint sets $A$ and $B$, such that every edge
$e\in E$ connects  one vertex in $A$ and one in $B$.
A length-$l$ {\it walk} in $G$ is a successive series of edges $e_i$ and vertices
$v_j$ such as $v_1e_1v_2e_2\cdots v_le_lv_{l+1}$, forming a continuous curve, i.e. each $e_i$ connects
$v_i$ to $v_{i+1}$. A walk is closed if
the initial and terminal vertices are the same, i.e $v_1=v_{l+1}$.
A closed walk in which only the end vertices are the same is called a {\it
cycle}, i.e. $v_{i}\ne v_{j}$ for each $1\le i< j\le l+1$ except for
$i=1$, $j={l+1}$. The {\it girth} of a graph is defined
as the length of its shortest cycles.

To a given  parity-check matrix $H$, a bipartite graph referred to as its {\it Tanner graph} (TG) is associated
 in which one set of vertices, called the check nodes, represent the set of rows of $H$, and the other
set, called the bit nodes (or symbol nodes) represent the set of  columns of $H$.
A symbol node is adjacent with a check node if and only if the entry of $H$ located in the corresponding row and
column is nonzero.

Let $m$ and $s$ be two nonnegative integers with $0\le s\le m-1$. The
$m\times m$ circulant permutation matrix ${\cal I}_m^s$ is the matrix obtained from $m\times m$ identity matrix
${\cal I}_m$ by shifting its rows $s$ positions to the left. It is clear that ${\cal I}_m^0={\cal
I}$. For simplicity, ${\cal I}_m^s$ is denoted by ${\cal I}^s$
when $m$ is known. It is noticed that this definition of circulant permutation matrix seems to be different with~\cite{struct4}, i.e. the row-shift is right-wise instead of left-wise, however it can be seen easily that the same conclusions can be obtained.
\medskip

The definition of the block-structure graph given in
\cite{cylinder} is reproduced as follows. Let $m$, $b$ and
$\gamma$ be  positive integers and $b<\gamma$. Let ${\cal
H}=(H_{i,j})_{b\times \gamma}$, wherein each $H_{i,j}$ is either a
$m\times m$ circulant permutation matrix or the $m\times m$
all-zero matrix. Considering $\cal H$ as a ${b\times \gamma}$
matrix with $m\times m$ entries, we refer to $\cal H$ as a matrix
having $b$ block-rows and $\gamma$ block-columns. For simplicity,
the matrix $\cal H$ and the QC-LDPC code represented by $\cal H$
as its parity-check matrix are referred to
 as $m-$circulant matrix and $m-$circulant code, respectively. The block-structure graph
associated to $\cal H$, denoted by BSG$({\cal H})$, is defined as follows.
\begin{defn}
\rm\label{def-bsg} Consider a set of vertices $V=\{v_1,v_2,...,v_b\}$, where $v_i$ represents the $i$th
block-row of $\cal H$. For each $i,j\in \{1,\cdots,b\}$ and $k\in
\{1,\cdots,\gamma\}$, where $H_{i,k}={\cal I}^{s_1}$ and
$H_{j,k}={\cal I}^{s_2}$ for some $0\le s_1, s_2\le m-1$, two
vertices $v_i,v_j\in V$ are joined by two directed edges labeled
with $(k, s)$, from $v_i$ to $v_j$, and $(k, s')$, from $v_j$ to
$v_i$, wherein $s = -s'=s_2 - s_1 \bmod m$. For each edge of the so constructed graph $G$,
the first and second component of its label $(k, s)$ are referred
to as the {\it column index} and the {\it shift} of that edge,
respectively. The resulting graph $G$ is called the {\it
block-structure graph} of $H$, and is denoted by BSG$({\cal H})$.
\end{defn}

\begin{defn}\rm
Consider a $m$-circulant matrix $\cal H$ and its associated BSG $G$.
A length-$l$ closed walk in $G$ is given by a sequence of vertices
$v_{i_1},v_{i_2},\cdots,v_{i_{l}},v_{i_{l+1}}$, where
$i_{l+1}=i_1$, with edges $e_1,e_2,...,e_{l}$ such that for each
$1\le j \le l$, the edge $e_j$, labeled with $(k_j,s_j)$, connects the vertices
$v_{i_j}$ and $v_{i_{j+1}}$ and that the following conditions hold:
\begin{enumerate}
\item Each edge $e_j$ in the sequence $e_1,e_2,...,e_{l}$
is repeated at most $m$ times; \item For each $1\le j \le l$,
$k_j\ne k_{j+1}$, where $k_{l+1}:=k_1$, i.e. the index columns of successive edges are different;  \item $\sum_{j=1}^{l}
s_j\equiv 0\pmod m$, i.e. the shift sum of the edges is zero modulo $m$.
\end{enumerate}
\end{defn}
For simplicity, we may show this length-$l$ closed
walk, or briefly  closed $l$-walk, in $G$ by the following chain:
$$ \begin{array}{c}
v_{i_1}\stackrel{(k_1,s_1)}{\longrightarrow}
v_{i_2}\stackrel{(k_2,s_2)}{\longrightarrow} v_{i_3}\quad...\quad
v_{i_{l-1}}\stackrel{(k_{l-1},s_{l-1})}{\longrightarrow}v_{i_l}\stackrel{(k_l,s_l)}{\longrightarrow}v_{i_1}.\end{array}$$

\begin{figure}[ht]
\begin{center}
\includegraphics[scale=0.85]{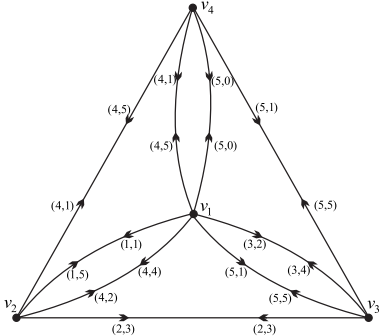}
\end{center}
\caption{\small The BSG of the 6-circulant code given in Example \ref{ex-bsg}.}
\label{bsg1}
\end{figure}

\begin{ex}
\label{ex-bsg} \rm
Consider the following 6-circulant matrix $\cal H$.
$${\cal H}=\left(\begin{array}{cccccc}{\cal I}&&{\cal I}&{\cal I}&{\cal I}\\
{\cal I}^1&{\cal I}&&{\cal I}^4&&\\
&{\cal I}^3&{\cal I}^2&&{\cal I}^1\\
&&&{\cal I}^5&{\cal I}
\end{array}\right)$$
Figure~\ref{bsg1} shows BSG$({\cal H})$, wherein each two-tuple
near an arrow, shows the label of that
edge from its initial vertex to its terminal vertex. The following
chain shows a closed $4$-walk in BSG$({\cal H})$ as
$1\ne4\ne1\ne4$ and $1+2+1+2=0\pmod 6$.
$$\begin{array}{c}v_1\stackrel{(1,1)}{\longrightarrow}v_2\stackrel{(4,2)}
{\longrightarrow}v_1\stackrel{(1,1)}{\longrightarrow}v_2\stackrel{(4,2)}{\longrightarrow}v_1.\end{array}$$
\end{ex}

The relationship between the length of the shortest closed walks
in BSG of a circulant code and the girth of its Tanner graph
is considered in the following Lemma~\cite{cylinder}.
\begin{lem}\rm
For each circulant matrix $\cal H$, the girth of
${\cal H}$, i.e. the length of the shortest cycle in TG$({\cal H})$,
is twice of the length of the shortest closed walk in BSG$({\cal H})$.
\end{lem}

In the sequel, the definition of a $(v,b,t)$-finite set system
(see \cite{handbook}) and some examples are presented.
\begin{defn}
\rm \label{fss} A $(v,b,t)$-finite set system, denoted by FSS$(v,
b,t)$, is a pair $(V,{\cal B})$ where $V$ is a finite set of size
$v$ whose members are called points, equipped with a finite
collection $\cal B$ of size $b$ consisting subsets of $V$ whose
members are called blocks, together with a positive integer $t$
less than or equal to the maximum block-size among the size of
blocks in $\cal B$. \\

According to Definition~\ref{fss}, a block $B$ in $\cal B$ may be
repeated, so, the word "collection" is preferred to "set". For a
given FSS$(v, b,t)$, we show its points and collection of blocks
by $V=\{1,2,\cdots,v\}$ and ${\cal B}=[B_1,B_2,\cdots,B_b]$,
respectively. For such a finite set system, we set the followings:
\begin{itemize}
\item $ K = [|B_1|,|B_2|,\ldots, |B_b|]$ is the collection of
block sizes;
\item For each $x\in V$, let $r_x$ be the replication
number of $x$ in the blocks of ${\cal B}$, i.e. the number of
blocks $B\in{\cal B}$ containing $x$ and let $R=[r_x: x\in V]$ be
the collection of replication numbers; \item For each $0\le i \le
t$ and each $i-$subset $T$ of $V$, let $\Lambda_{i,T}$ be the
number of blocks $B\in{\cal B}$ containing $T$ and let $\Lambda_i
=\{\Lambda_{i,T} : T \subseteq V, |T|=i\}$. Therefore, $\Lambda_0
= \{b\}$
and 
$\Lambda_1$ is the set of distinct elements in $R$.
\end{itemize}
\end{defn}

For a given FSS$(v, b,t)$, a set of distinct points
$x_1,x_2,\ldots,x_t$ in $V$ are called {\it co-block} if there
exists at least one block $B_i\in{\cal B}$ such that
$x_1,\ldots,x_t\in B$.

\begin{table}
\begin{center}
$\begin{array}{|cccccc|}
\hline
K&R&\Lambda_2&\Lambda_t&{\rm Other}&{\rm Name}\\
\hline\hline
{[k,\ldots,k]}&{[r,\ldots,r]}&\{\lambda_2\}&\{\lambda_t\}&&t-\hbox{design}\\
{[k,\ldots,k]}&{[r,\ldots,r]}&\{\lambda_2\}&\{\lambda_t\}&\lambda_t=1&\hbox{Steiner system}\\
{[k,\ldots,k]}&{[r,\ldots,r]}&\{\lambda\}&&&\hbox{balanced incomplete block design (BIBD)}\\
&{[r,\ldots,r]}&\{\lambda\}&&&(r,\lambda)-\hbox{design}\\
&&\{\lambda\}&&&\hbox{pairwise balanced design (PBD)}\\
&&\{\lambda\}&&\lambda=1&\hbox{linear space}\\
&&&\{\lambda_t\}&&t-\hbox{wise balanced design}\\
\hline
\end{array}$
\caption{\footnotesize Some of the set systems that result when at
least one of the balance sets  $\{\Lambda_t\}$ is a
singleton.\label{tab-FSSaa}}
\end{center}
\end{table}

\begin{ex}\label{fsse}\rm
\label{ex-fss} The following 10 blocks define a FSS$(8,10,3)$:~~
$B_{1}=\{1,2,4,5\}$, $B_{2}=\{1,2,3,7\}$, $B_{3}=\{1,3,5,8\}$,
$B_{4}=\{2,3,5,6\}$, $B_{5}=\{2,3,4,8\}$, $B_{6}=\{3,4,6,7\}$,
$B_{7}=\{4,5,7,8\}$, $B_{8}=\{1,4,6,8\}$, $B_{9}=\{1,5,6,7\}$,
$B_{10}=\{2,6,7,8\}$. For this, $K=[3,3,\ldots,3]$,
$R=[5,5,\ldots,5]$, $\Lambda_0=\{10\}$, $\Lambda_1=\{5\}$,
$\Lambda_2=\{2\}$ and $\Lambda_3=\{0,1\}$.
\end{ex}

Setting restrictions on $\{\Lambda_t\}$s imposes specific
structures on the system. Restriction to a single element balances
the structure in some regard. Table~\ref{tab-FSSaa} considers the
effect of restricting these sets to singleton sets. Indeed, the
set of so obtained systems are among the most central objects in
design theory and have been studied extensively. If in a
FSS$(v,b,t)$ we have $K=[k,\cdots,k]$, $R=[r,\cdots,r]$,
$\Lambda_t=\{\lambda_t\}$, then the system is called $t$-design.
Moreover, if $\lambda_t=1$, then the $t$-design is called Steiner
system $S(t,k,v)$. A {\it balanced incomplete block design}
(BIBD$(v,k,\lambda)$) is a $2$-design with $\lambda=\lambda_2$.
For example, the FFS explained in example \ref{fsse} with bloks
${\cal B}=[B_1,B_2,\ldots,B_{10}]$ can be considered as an
BIBD$(8,4,2)$. A {\it pairwise balanced design} (PBD) is a FSS$(v,
b,t)$ with $\Lambda_2=\{\lambda\}$.
\\

The incidence matrix of a given FSS is defined in the following.

\begin{defn}
\label{t-code}\rm Let $(V,{\cal B})$ be a FSS$(v,b,t)$ where
${\cal B}=[B_1,B_2,\cdots,B_b]$. Let $\{T_1,T_2,\cdots,T_\gamma\}$
be the set of all $(t-1)$-subsets of $V$ such that for each $T_i$
there are at least two distinct blocks  containing $T_i$. We
assign a $(0,1)$-matrix $H=(h_{i,j})_{b\times \gamma}$ to
$(V,{\cal B})$ wherein  the $i$-th row ($j$-th column) of $H$
represents $B_i$ (resp. $T_j$), and $h_{i,j}=1$ if and only if
$T_j\subseteq B_i$. The matrix $H$ is referred to as the {\it
incidence matrix} of $(V,{\cal B})$.
\end{defn}

For example, the incidence matrix $H$ associated with the system
defined in Example~\ref{ex-fss} is given by (\ref{mat-base}).

\begin{equation}
\label{mat-base}  {\small H=}\begin{array}{rl}
&\hspace{-4mm}{\tiny
\begin{array}{cccccccccccccccccccccccccccc}
1,\hspace{-0.5mm}2\hspace{1.5mm}1,\hspace{-0.5mm}3\hspace{1.5mm}1,\hspace{-0.5mm}4\hspace{1.5mm}
1,\hspace{-0.5mm}5\hspace{1.5mm}1,\hspace{-0.5mm}6\hspace{1.5mm}1,\hspace{-0.5mm}7\hspace{1.5mm}
1,\hspace{-0.5mm}8\hspace{1.5mm}2,\hspace{-0.5mm}3\hspace{1.5mm}2,\hspace{-0.5mm}4\hspace{1.5mm}
2,\hspace{-0.5mm}5\hspace{1.5mm}2,\hspace{-0.5mm}6\hspace{1.5mm}2,\hspace{-0.5mm}7\hspace{1.5mm}
2,\hspace{-0.5mm}8\hspace{1.5mm}3,\hspace{-0.5mm}4\hspace{1.5mm}3,\hspace{-0.5mm}5\hspace{1.5mm}
3,\hspace{-0.5mm}6\hspace{1.5mm}3,\hspace{-0.5mm}7\hspace{1.5mm}3,\hspace{-0.5mm}8\hspace{1.5mm}
4,\hspace{-0.5mm}5\hspace{1.5mm}4,\hspace{-0.5mm}6\hspace{1.5mm}4,\hspace{-0.5mm}7\hspace{1.5mm}
4,\hspace{-0.5mm}8\hspace{1.5mm}5,\hspace{-0.5mm}6\hspace{1.5mm}5,\hspace{-0.5mm}7\hspace{1.5mm}
5,\hspace{-0.5mm}8\hspace{1.5mm}6,\hspace{-0.5mm}7\hspace{1.5mm}6,\hspace{-0.5mm}8\hspace{1.5mm}7,\hspace{-0.5mm}8
\end{array}}\\\tiny
\begin{array}{l}B_1\\B_2\\B_3\\B_4\\B_5\\B_6\\B_7\\B_8\\B_9\\B_{10}\end{array}\tiny
&\hspace{-6mm}\tiny\left(\begin{array}{cccccccccccccccccccccccccccc}
 1&0&1&1&0&0&0&0&1
&1&0&0&0&0&0&0&0&0&1&0&0&0&0&0&0&0&0&0\\ 1&1&0&0&0&1
&0&1&0&0&0&1&0&0&0&0&1&0&0&0&0&0&0&0&0&0&0&0\\ 0&1&0
&1&0&0&1&0&0&0&0&0&0&0&1&0&0&1&0&0&0&0&0&0&1&0&0&0
\\ 0&0&0&0&0&0&0&1&0&1&1&0&0&0&1&1&0&0&0&0&0&0&1&0&0
&0&0&0\\ 0&0&0&0&0&0&0&1&1&0&0&0&1&1&0&0&0&1&0&0&0&1
&0&0&0&0&0&0\\ 0&0&0&0&0&0&0&0&0&0&0&0&0&1&0&1&1&0&0
&1&1&0&0&0&0&1&0&0\\ 0&0&0&0&0&0&0&0&0&0&0&0&0&0&0&0
&0&0&1&0&1&1&0&1&1&0&0&1\\ 0&0&1&0&1&0&1&0&0&0&0&0&0
&0&0&0&0&0&0&1&0&1&0&0&0&0&1&0\\ 0&0&0&1&1&1&0&0&0&0
&0&0&0&0&0&0&0&0&0&0&0&0&1&1&0&1&0&0\\ 0&0&0&0&0&0&0
&0&0&0&1&1&1&0&0&0&0&0&0&0&0&0&0&0&0&1&1&1
\end{array}\right)\end{array}
\end{equation}


In definition \ref{t-code}, without loss of generality, we may
assume that $t=2$, as it is clear that the incidence matrix of a
FSS$(v,b,t)$ can be considered as the incidence matrix of a
FSS$(v',b,2)$, where $v'\le{v\choose t-1}$. Hence in the rest of
the paper, by FSS we mean a FSS$(v,b,2)$ for some positive
integers $v$ and $b$. In this case, $\gamma$ is equal to $v$.

\begin{ex}\rm
The incidence matrix of the FSS$(8,10,3)$, represented by
(\ref{mat-base}), can be considered as the base matrix of a
FSS$(28,10,2)$, $(V,{\cal B})$, with $V=\{1,2,\cdots,28\}$ and
${\cal B}=[B_1,B_2,\cdots,B_{10}]$, for $B_1=\{1, 3, 4, 9, 10,
19\}$, $B_{2}=\{1, 2, 6, 8, 12, 17\},$ $B_{3}=\{2, 4, 7, 15, 18,
25\},$ $B_{4}=\{8, 10, 11, 15, 16, 23\},$ $B_{5}=\{8, 9, 13, 14,
18, 22\},$ $B_{6}=\{14, 16, 17, 20, 21, 26\},$ $B_{7}=\{19, 21,
22, 24, 25, 28\},$ $B_{8}=\{3, 5, 7, 20, 22, 27\},$  $B_{9}=\{4,
5, 6, 23, 24, 26\},$ $B_{10}=\{11, 12, 13, 26, 27, 28\}$.
\end{ex}
\begin{defn}\rm
Let $(V,{\cal B})$ be a FSS$(v,b,2)$ where ${\cal
B}=[B_1,B_2,\cdots,B_b]$  and let $m$ be a positive integer. To the element $i\in B_j$ a non-negative
integer $s_{i,j}$ is assigned. A finite
sequence $S=(s_{i,j})_{i\in B_j,1\le j\le b}$ on $\cal B$ is
called a {\it shift sequence} of order $m$ if each $s_{i,j}$
is in $\Bbb{Z}_m$.

\medskip
Let $S=(s_{i,j})_{i\in B_j,1\le j\le b}$ be a shift sequence of
order $m$ on ${\cal B}$, and ${\cal H}=(H_{i,j})_{v\times
b}$ be the matrix consisting of $m\times m$ block circulant matrices
$H_{i,j}$, where:
\begin{equation}
\label{pc}\small H_{i,j}=\left\{\begin{array}{cc}{\cal
I}^{s_{i,j}}&i\in B_j,\\0&i\not\in B_j.\end{array}\right.
\end{equation}
A finite set system code (FSS-code) based on $\cal B$ and $S$, denoted by $C_{m,{\cal B},S}$, is defined as the quasi-cyclic LDPC code with the parity-check matrix $\cal H$ if $b < {v\choose t-1}$ or its transpose ${\cal H}^T$, elsewhere. Hereinafter, without loss of generality, we assume that $b > {v\choose t-1}$.


\end{defn}
\begin{ex}\rm
\label{632}
Let $V=\{1,2,\cdots,6\}$ and ${\cal B}=[\{1, 2, 5\}$,
$\{1, 2, 6\},$ $\{1, 3, 4\}$, $\{1, 3, 5\}$, $\{1, 4, 6\}$, $\{2,
3, 4\}$, $\{2, 3, 6\}$, $\{2, 4, 5\}$, $\{3, 5, 6\}$, $\{4, 5,
6\}]$. Then $(V,{\cal B})$, known as BIBD(6,3,2), is a
FSS$(6,10,2)$.

Now let  $S=(s_{i,j})_{i\in B_j,1\le j\le 10}$ be a shift sequence
on $(V,{\cal B})$ and $\cal H$ be the associated parity-check matrix.
This code  has rate at least $1-6/10=0.4$. Without loss of generality, up to code equivalence,
we may assume that the first shift $s_{i,j}$ for each column of $\cal H$ is zero, i.e. $s_{1,1}=
s_{1,2}= s_{1,3}= s_{1,4}=
  s_{1,5}= s_{2,6}= s_{2,7}= s_{2,8}= s_{3,9}= s_{4,10}= 0$.
 Thus $${\cal H}=\left(\begin{array}{cc}
\begin{array}{ccccc}
       {\cal I} & {\cal I} & {\cal I} & {\cal I} & {\cal I}  \\
       {\cal I}^{s_{2,1}} & {\cal I}^{s_{2,2}} & 0 & 0 &0\\
      0  & 0 & {\cal I}^{s_{3,3}} & {\cal I}^{s_{3,4}} &0 \\
      0  & 0 & {\cal I}^{s_{4,3}} & 0 &  {\cal I}^{s_{4,5}}\\
       {\cal I}^{s_{5,1}} & 0 & 0 & {\cal I}^{s_{5,4}} &0 \\
       0 & {\cal I}^{s_{6,2}} & 0 & 0 &  {\cal I}^{s_{6,5}}\\
     \end{array}
&\hspace{-.5cm}\begin{array}{cccccccccc}
       0 & 0 & 0 & 0 & 0 \\
        {\cal I} & {\cal I} & {\cal I} & 0 & 0 \\
        {\cal I}^{s_{3,6}} & {\cal I}^{s_{3,7}} & 0 & {\cal I} & 0 \\
        {\cal I}^{s_{4,6}} & 0 & {\cal I}^{s_{4,8}} & 0 & {\cal I} \\
      0 & 0 & {\cal I}^{s_{5,8}} & {\cal I}^{s_{5,9}} & {\cal I}^{s_{5,10}} \\
      0  & {\cal I}^{s_{6,7}} & 0 & {\cal I}^{s_{6,9}} & {\cal I}^{s_{6,10}} \\
     \end{array}
  \end{array}\right).$$
\end{ex}

The numbers $R$ and $K$ in definition \ref{fss} are the row and
column-weight distributions of the incidence matrix $H$,
respectively. Hence, by choosing a FSS$(v,b,2)$ with desired
numbers $R$ and $K$ we can construct parity-check matrices having
suitable column-weight and row-weight distributions.
 On the other hand, the rate of a FSS-code $C_{m,{\cal B},S}$ is at
least $1-\frac{{\rm min} \{b , v \}}{{\rm max}\{b,v
\}}$. Hence, by increasing $|v-b|$, we hope that the obtained rate will be
large enough. In order to have the girth of ${\cal H}$ large enough, we
first note that the girth of $C_{m,{\cal B},S}$ represented by ${\cal H}$ is
upper-bounded by $g({\cal B})$, where $g({\cal B})$ is the largest achievable girth
for all possible integers $m$ and shift-sequences $S$ on $m$.

Let $H$ be the incidence-matrix of $(V,{\cal B})$. Using
Proto-graphs~\cite{protograph}, Kim et. al.~\cite{sun} have
introduced some incidence matrices, denoted $P_{2i}$, and have
claimed that if $H$ does not contain $P_{2i}$ and $P_{2i}^T$ for
all $i< g$, then $g({\cal B})$ will be at least $2g$. But,
checking this condition, i.e. ensuring that $P_{2i}$ and
$P_{2i}^T$, $i<g$, are not as sub-matrices in $H$, especially when
$g$ is large, is a process with high computational complexity as
$H$ can include any row (column) permutation of $P_{2i}$ or
$P_{2i}^T$, as its sub-matrices.

In this paper, by introducing some special chains in $(V,{\cal B})$, called inevitable walks,
we propose a theorem that efficiently determines $g({\cal B})$. Then, we propose two efficient algorithms
for finding $(V,{\cal B})$ with a desired girth $g({\cal B})$. Finally, we propose a modified version of the code-generating algorithm given in \cite{ghol4}, such that for a given $(V,{\cal B})$  the algorithm
efficiently generates FSS-codes with a desired girth.  First, we define inevitable walks in a FSS as they have an important role in finding the maximum achievable girth by FSS-codes based on a given mother matrix.

\begin{figure}[ht]
 \begin{center}
\includegraphics[scale=0.5]{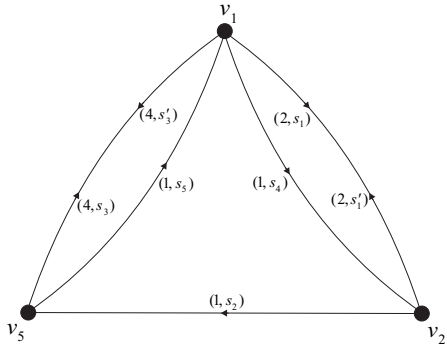}
\end{center}
\caption{\small The length-7 closed walk associated with the length-7 inevitable walk in BSG of Example \ref{ex-walk-7}.} \label{walk-7}
\end{figure}

\begin{defn}\rm
Let $V=\{1,2,\cdots,v\}$ and ${\cal B}=[B_1,B_2,\cdots,B_b]$ form a
FSS$(v,b,2)$. An inevitable walk of length $\ell$, briefly
$\ell$-inevitable walk, in $\cal B$ consists of a list of integers
$i_1,i_2,\cdots,i_{\ell }\in V$ and
$k_1,k_2,\cdots,k_{\ell }\in\{1,2,\cdots,b\}$, denoted by
$[i_1,k_1,i_2,k_2,\cdots,i_{\ell },k_{\ell }]$, such that the
following conditions hold wherein $i_{\ell +1}=i_1, k_{\ell +1}=k_1,
i_{\ell +2}=i_2, k_{\ell +2}=k_2,\cdots$):
\begin{enumerate}
\item $i_j\ne i_{j+1}$, $i_j,i_{j+1}\in B_{k_j}$ for $1\le
j\le \ell $.
\item For $1\le j\le \ell $, if $k_j=k_{j+1}$ then
$i_j\ne i_{j+2}$.
\item For some $p$, the vectors
$(i_1,k_1,i_2),(i_2,k_2,i_3),\cdots,(i_{\ell -1},k_{\ell -1},i_{\ell}),(i_{\ell },k_{\ell },i_1)$ can be
partitioned into $p$ sets of the form
$A_j=\{(i_{j_1},k_{j},i_{j_2}),(i_{j_2},k_{j},i_{j_3}),\cdots
,(i_{j_{t-1}},k_{j},i_{j_t}),(i_{j_t},k_{j},i_{j_1})\}$, $1\le j\le p$, where $t$ depends on $j$.
\end{enumerate}
\end{defn}
Let $G$ be the BSG of a parity-check matrix of a given FSS$(m,{\cal B},S)$ code. If
$[i_1,$ $k_1,$ $i_2,$ $k_2,$ $\cdots,$ $i_{\ell },$ $k_{\ell }]$ is an $\ell $-inevitable walk in
$\cal B$, then the following $\ell $-closed walk always exists in $G$,
regardless of the values of $m$ and $S$.
$$\begin{array}{c}v_{i_1}\stackrel{(k_1,s_1)}{\longrightarrow} v_{i_2}\stackrel{(k_2,s_2)}{\longrightarrow} v_{i_3}\quad...\quad
v_{i_{\ell-1}}\stackrel{(k_{\ell-1},s_{\ell-1})}{\longrightarrow}v_{i_\ell}\stackrel{(k_\ell, s_\ell)}{\longrightarrow}v_{i_1},\end{array}$$
where $s_j=s_{i_{j+1},k_j}-s_{i_{j},k_j}\bmod m$ for $i_j,i_{j+1}\in B_{k_j}$.
\begin{ex}\rm
\label{ex-walk-7}
Let $(V,{\cal B})$ be the ${\rm FSS}(6,10,2)$ given by example
\ref{632}. The sequence [1, 2, 2, 1, 5, 4, 1, 1, 2, 2, 1, 4, 5, 1] is a
7-inevitable walk in $\cal B$. In fact, the vectors
(1,2,2), (2,1,5), (5,4,1), (1,1,2), (2,2,1), (1,4,5), and (5,1,1) can
be partitioned to the sets $A_1=\{(1,2,2),(2,2,1)\}$,
$A_2=\{(1,1,2),(2,1,5),(5,1,1)\}$, and $A_3=\{(5,4,1),(1,4,5)\}$.
This inevitable walk is represented by the following chain:
$$\begin{array}{c}v_{1}\stackrel{(2,s_1)}{\longrightarrow} v_{2}\stackrel{(1,s_2)}
{\longrightarrow} v_{5}\stackrel{(4,s_3)}{\longrightarrow}
v_{1}\stackrel{(1,s_4)} {\longrightarrow}
v_{2}\stackrel{(2,s'_1)}{\longrightarrow} v_{1}\stackrel{(4,s'_3)}
{\longrightarrow} v_{5}\stackrel{(1,s_5)}{\longrightarrow}
v_{1},\end{array}$$ where $s_1=-s'_1=s_{2,2}-s_{1,2}$,
$s_2=s_{5,1}-s_{2,1}$, $s_3=-s'_3=s_{1,4}-s_{5,4}$,
$s_4=s_{2,1}-s_{1,1}$ and $s_5=s_{1,1}- s_{5,1}$.
It is clear that the sum of the shift of consecutive edges of this chain,
shown by a 7-inevitable walk in Figure~\ref{walk-7}, is zero modulus $m$, for any $m$, i.e.
\[
\begin{array}{l}s_1+ s_2+ s_3+ s_4+
s'_1+ s'_3+ s_5=\\(s_{2,2}-s_{1,2})+ (s_{5,1}-s_{2,1})+
(s_{1,4}-s_{5,4})+ (s_{2,1}-s_{1,1})\\+ (s_{1,2}-s_{2,2})+
(s_{5,4}-s_{1,4})+ (s_{1,1}- s_{5,1})=0\pmod m.\end{array}\]
\end{ex}

Let $(V,{\cal B})$  be a FSS$(v,b,2)$ and $H=(H_{i,j})_{b\times
v}$ be its associated incidence matrix. Then, an $\ell$-inevitable
walk in $\cal B$ is a closed walk that exists in the BSG of
FSS$(m,{\cal B},S)$ codes regardless of the values of $m$ and $S$.

The upper-bound $g({\cal B})$ is obtained from the shortest inevitable walks in $\cal B$ by the following theorem.
\begin{thm}\rm
\label{thm1}
The number $g({\cal B})$ is the smallest integer $2\ell$, such that there
is at least one $\ell$-inevitable walk in $\cal B$.\\
\pf We should show that for a given $(V,{\cal B})$, the maximum achievable girth of FSS$(m,{\cal B},S)$ codes,
for all $m$ and $S$, is the smallest $2\ell$, such that there is an $\ell$-inevitable walk in $\cal B$.
Let $W=[i_1, k_1, \cdots, i_{\ell}, k_{\ell}]$ be an inevitable walk in $\cal B$ and that $\ell$ is the smallest such number.
We show that the girth of any FSS$(m,{\cal B},S)$ code is upper-bounded by $2\ell$. If $\cal H$ is the parity-check matrix of a
FSS$(m,{\cal B},S)$ code, for some $m$ and $S$, and $G$ is the BSG$({\cal H})$, then regardless of the values of $m$ and $S$, $G$ contains
the $\ell$-closed walk represented by the following chain.
\begin{equation}\label{ch4}{\begin{array}{c}
v_{i_1}\stackrel{(k_1,s_1)}{\longrightarrow}v_{i_2}\stackrel{(k_2,s_2)}{\longrightarrow}
v_{i_2}\quad...\quad
v_{i_{\ell}}\stackrel{(k_{\ell},s_{\ell})}{\longrightarrow}v_{i_1},\end{array}}\end{equation}
where $s_j=s_{i_{j+1},k_j}-s_{i_{j},k_j}\bmod m$, $1\le j\le
\ell$. As $\cal C$ is an inevitable walk, the vectors $(i_1,k_1,i_2)$, $(i_2,k_2,i_3)$, $\cdots$,
$(i_{\ell},k_{\ell},i_1)$ can be partitioned to the $p$ sets $A_j=$ $\{(i_{j_1},$ $k_{j}, i_{j_2}),$ $(i_{j_2}, k_{j}, i_{j_3}),$ $\cdots,$ $(i_{j_t}, k_{j}, i_{j_1})\}$, $1\le j\le p$, where $t=|A_j|$ and the points $i_{j_1},i_{j_2},\cdots,i_{j_t}$ are $B_{k_j}$ co-block. Hence, $\sum\limits_{h=1}^{|A_j|}(s_{i_{j_{h+1}},k_j}-s_{i_{j_{h}},k_j})\equiv0\bmod m$ for any $1\le j\le p$, and $\sum_{i=1}^{\ell}s_i=\sum\limits_{j=1}^{p}\sum\limits_{h=1}^{|A_j|}
(s_{i_{j_{h+1}},k_j}-s_{i_{j_{h}},k_j})\equiv 0\bmod m$ where $i_{j_{|A_j|+1}}=i_{j_1}$.

Conversely, if the $\ell$-closed walk, represented by the chain given by (\ref{ch4}), is in $G$, regardless of the values of $m$ and $S$,
then it is easily verified that each edge of this walk is a part of an $\ell$-closed walk
in $G$, between some points $i_{j_1},i_{j_2},\cdots,i_{j_t}$ belonging to a block $B_{k_j}$, for some $1\le j\le p$. Now, let
$A_j=\{(i_{j_1},k_{j},i_{j_2})$, $(i_{j_2},k_{j},i_{j_3})$,
$\ldots$, $(i_{j_t},k_{j},i_{j_1})\}$. It is clear that $W=[i_1$,
$k_1$, $\cdots$, $i_{\ell}$, $k_{\ell}]$ is an $\ell$-inevitable walk
in $\cal B$. \eep
\end{thm}

\section{Construction of FSS Codes with Arbitrary Girth}
By Theorem~\ref{thm1}, the number $g({\cal B})$ is the maximum achievable girth by FSS$(m,{\cal B},S)$ codes, for all $m$ and $S$,
and it is equal to $2\ell$, where $\ell$ is the smallest number such that ${\cal B}$ contains an $\ell$-inevitable walk.
Thus, in order to construct FSS codes with large girths, we first need to design ${\cal B}$ having $g({\cal
B})$ large enough, and then choose $m$ and $S$ properly, such that the associated FSS$(m,{\cal B},S)$ code has the desired
girth not greater than $g({\cal B})$.

We introduce two structured methods to construct $(V,{\cal B})$ with desired $g({\cal B})$.
 The first approach uses a recursive method to convert a FSS $(V',{\cal B}')$ with $g({\cal B}')=2g$ to a new FSS
$(V,{\cal B})$ with $g({\cal B})\ge 6g$. For given positive
integers $b$ and $v$ and a collection of integers $K$, the second
approach produces $(V,{\cal B})$, if it exists, such that its
incidence matrix $H=(h_{i,j})_{v\times b}$ has column weight
distribution $K$ and satisfies $g({\cal B})\ge 2g$.
\subsection{Method 1}
Let $(V',{\cal B}')$ be a FSS with $g({\cal B}')=2g'$ and $\cal H$
be the parity-check matrix of a FSS$(m,{\cal B}',S)$ code, for
some $m$ and $S$. If we consider $\cal H$ as the incidence matrix
of a new FSS, $(V,{\cal B})$, then Theorem 2 of \cite{sun}
guarantees that $g({\cal B})\ge 6g'$. Hence, to design $(V,{\cal
B})$ with $g({\cal B})\ge 2g$, we use the following algorithm.
\medskip

{\bf Algorithm I}
\begin{enumerate}
\item Design $(V',{\cal B}')$, such that $3g({\cal B}')\ge 2g$.
\item Choose a positive integer $m$ and a shift-sequence $S$ on
$m$ properly, such that their corresponding FSS$(m,{\cal B}',S)$
code, with the parity check matrix $\cal H$, has girth $2g'$ ($\le
g({\cal B}')$), where $3g'\ge g$. \item Consider $\cal H$ as the
incidence matrix of a new FSS $(V,{\cal B})$ where $g({\cal B})\ge
6g'\ge 2g$.
\end{enumerate}
By continuing this process, one can obtain FSS codes with arbitrary
large girths. Based on Algorithm I, Table~\ref{tab-struc1} presents some new $(V,{\cal
B})$ with $g({\cal B})=2g$, derived from the initial
$(V',{\cal B}')$ with $g({\cal B}')=2g'$ where $g'=\lceil g/3\rceil$.
\begin{rem}\rm
We should note that the row (column) weight distribution of the
incidence matrices of $(V,{\cal B})$  and $(V',{\cal B}')$ in
Method 1 are the same. Now, in order to construct $(V,{\cal B})$
with a incidence matrix having a desired column-weight
distribution, we use the following construction method.
\end{rem}
\subsection{Method 2}
Let $b$, $g$ and $v$ be some positive integers and
$K=[k_1,\cdots,k_b]$ be a collection of not necessarily distinct
positive integers $k_i$. In the sequel, we provide an algorithm
which inductively generates proper FSS$(v,b,2)$ whose incidence
matrices can be considered as the incidence matrices of some QC
LDPC codes with column-weight distribution $K$ and girth at least
$2g$. Moreover, we assume $k_i>1$ to avoid having column-weight 1
in the associated parity-check matrix $\cal H$. In order to
construct a $(V,{\cal B})$ having incidence matrix with
column-weight distribution $K$, we propose the following
deterministic algorithm which inductively finds ${\cal
B}=[B_1,B_2,\cdots,B_b]$, $|B_i|=k_i$, $B_i\subseteq
V=\{1,\cdots,v\}$ with $g({\cal B})\ge 2g$.

An overview of the algorithm is as follows. Given a column-weight distribution $K=[k_1,\cdots,k_b]$ and girth $2g$, we form the elements of the required $\cal B$ sequentially from the elements of $B_1$ to $B_b$. In step $e$, assuming that the blocks $[B_1,\ldots,B_{b'-1}, B'_{b'}]$, $b'\le b$ and $B'_{b'}\subseteq B_{b'}$ are chosen, we determine $A_{e+1}$ consisting of all elements of $\beta \in V$, such that for any element $\beta_{b',j}\in B'_{b'}$, the inevitable walks containg the edge $(\beta_{b',j},\beta)$ in ${\cal B}_{e+1}=[B_1,\ldots,B'_{b'-1},B'_{b'}\cup\{\beta\}]$ have length at least $2g$, i.e. $g((\beta_{b',j},\beta),{\cal
B}_{e+1})\ge 2g$. Also, in order to avoid checking edges repeatedly, we have
assumed that $\beta>\max B'_{b'}$. If $A_{e+1} =\emptyset$,  then we move one step back and replace $A_e$ with $A_e-\{\beta_{b',|B'_{b'}|}\}$, choose an arbitrary element of the updated set $A_e$ as $\beta_{b',|B'_{b'}|+1}$, update $B'_{b'}=B'_{b'}\cup \{\beta_{b',|B'_{b'}|+1}\}$ and ${\cal
B}_{e}=[B_1,\cdots,B'_{b'}]$, then determine $A_{e+1}$. If this process ends up with $A_1 =\emptyset$  we conclude that there is no $(V,{\cal B})$ with $g({\cal B})=2g$. The formal step-by-step framework of the algorithm is as follows.

\medskip

{\bf Algorithm II}
\begin{description}
\item{s1.} Set $e=1$, $V=\{1,2,\cdots,v\}$, $b'=1$ and $A_1:=V$.
\item{s2.} First choose $\beta_{1,1}\in A_1$ and set $B_1=\{\beta_{1,1}\}$ and ${\cal B}_1=[B_1]$.
\item{s3.} In step $e$, assume that ${\cal B}_{e}=[B_1,\cdots,B'_{b'}]$, $1\le b'\le b$, has been constructed, such
that $|B_i|=k_i$, for each $i<b'$, and $|B'_{b'}|\le k_{b'}$.
\item{s4.} In step $e+1$, if
$|B'_{b'}|<k_{b'}$, then let $A_{e+1}$ be the set of all $\beta\in V$, such
that $\beta>\max B'_{b'}$ and $g((\beta_{b',j},\beta),{\cal
B}_{e+1})\ge 2g$, for all $j\le |B'_{b'}|$ where ${\cal
B}_{e+1}=[B_1,\cdots,B'_{b'-1},B'_{b'}\cup\{\beta\}]$, otherwise  if
$|B'_{b'}|=k_{b'}$ then define $A_{e+1}:=V$, replace $b'$ with $b'+1$ ($b'\to b'+1$), choose $\beta\in V$ and set
$B'_{b'}=\{\beta\}$, and go to s6.
\item{s5.} If $A_{e+1}=\emptyset$,
then $A_e\to A_e-\{\beta_{b',|B'_{b'}|}\}$, $e\to e-1$.
\item{s6.} If $e<0$, then there is no solution, so go to end.
\item{s7.} If $A_{e+1}\ne\emptyset$, then choose $\beta_{b',|B'_{b'}|+1}\in
A_{e+1}$, and update $B'_{b'}=B'_{b'}\cup \{\beta_{b',|B'_{b'}|+1}\}$, ${\cal
B}_{e+1}=[B_1,\cdots,B'_{b'}]$ and $e\to e+1$; else go to s5.
\item{s8.} If $e<\sum_{i=1}^{b} k_i$, then go to s4; else print ${\cal B}={\cal B}_e$ as a solution and go
to end.
\end{description}

For simplicity of the notations, we considered the convention that
$B_{i}=\{\beta_{i,1},\cdots,\beta_{i,k_i}\}$, $1\le i\le
b$, and that ${\cal B}_e=[B_1,\cdots,B_{b'-1},B'_{b'}]$ ($b'\le b$ and $B'_{b'}\subseteq B_{b'}$) is the sub-FSS of $\cal B$,
constructed in the first $e$ stages of the algorithm.
In fact, ${\cal B}_1=[\{\beta_{1,1}\}]$, ${\cal B}_2=[\{\beta_{1,1},\beta_{1,2}\}]$, $\ldots$, ${\cal B}_{k_1}=[B_1]$, ${\cal B}_{k_1+1}=[B_1,\{\beta_{2,1}\}]$, ${\cal B}_{k_1+2}=[B_1,\{\beta_{2,1},\beta_{2,2}\}]$, $\ldots$, ${\cal B}_{k_1+k_2}=[B_1,B_2]$, $\ldots$, ${\cal B}_{k_1+\cdots+k_b}={\cal B}$.
Also, $g((i_1,i_2),{\cal B}_{e})$, for some $i_1\in B'_{b'}$ and $i_2\in V$, is
the smallest integer number $2\ell$, such that there is an $\ell$-inevitable walk
$[i_1,j_1,i_2,j_2,\cdots,i_{\ell},j_1]$ in ${\cal B}_{e+1}$, where ${\cal B}_{e+1}=[B_1,\cdots,B'_{b'}\cup\{i_2\}]$ if $|B'_{b'}|<|B_{b'}|$, otherwise ${\cal B}_{e+1}=[B_1,\cdots,B_{b'},\{i_2\}]$. If there is
no such $\ell$-inevitable walk in ${\cal B}_{e+1}$, we set $g((i_1,i_2),{\cal B}_{e})=2g$ as condition $g({\cal
B})\ge 2g$ must be satisfied. Note that the maximum element in $B\subseteq V$ is denoted by $\max B$.\\

%

Applying algorithm II, we have constructed a set of $(V,{\cal B})$ with $14\le g({\cal B})\le20$ and rates $r=1-\frac{v}{b}$, with parity-check matrices having column-weights varying among 2, 3 and 4 (see~Table \ref{tab-struc2-1}). Moreover, as an advantage of algorithm II, Table~\ref{tab-struc2-2} presents some $(V,{\cal B})$ which have column-weight two mother matrices, satisfy $g({\cal B})\ge24$, and have rates $r$ better than the rates found in~\cite{ghol2} and \cite{ghol4}. Also, the algorithms outputs show that the rates of such column-weight two FSS codes grow with the row-number of their mother matrices.
\subsection{Construction of FSS Codes}
\label{secalg}
Given a positive integer $g$, let $(V,\cal B)$ be a FSS
with $g({\cal B})=2g$, where ${\cal B}=[B_1,\cdots,B_b]$. Here, we
provide a deterministic algorithm that generates FSS codes with girth
$2g'(\le 2g)$. This algorithm is a modified version of the code-generating algorithm given in~\cite{ghol4}. To determine the
shift sequence $S=(s_{i,j})_{i\in B_j,1\le j\le b}$ on $\cal B$, for a sufficiently large block size $m$,
initially we consider $S$ as the length-$\sum_{i=1}^bk_i$ vector $S:=[s_{1,1}$, $\ldots,$ $s_{k_1,1},$ $s_{1,2},$ $\ldots,$
$s_{k_2,2},$ $\ldots,$ $s_{1,b},$ $\ldots,$ $s_{k_b,b}]$, where
$k_i=|B_i|$. Assuming that the first $e$ elements of $S$, denoted by
$S_e := [s_1, s_2, \cdots, s_e]$, are chosen, we determine
$A_{e+1}$ consisting of all elements of $V$ such that for any
element $s\in A_{e+1}$ the FSS$(m,{\cal B}_{e+1},S_{e+1}=[s_1, s_2, \cdots, s_e, s_{e+1}:=s])$ code has girth at least $2g'$.
If $A_{e+1}=\emptyset$, then we move one step back and replace
$A_e$ with $A_e-\{s_e\}$. If this process ends up with $A_1 =
\emptyset$, we conclude that for the given $m$ and $(V,{\cal B})$,
there is no FSS$(m,{\cal B},S)$ code with girth
$2g'$. This algorithm efficiently generates FSS
codes with girth at most 20. A list of so constructed codes are given in Table \ref{tab-code}.
In the following, we find the complexity of the proposed algorithm generating FSS codes with girth $g$.\\
\medskip

\subsection{Complexity of the Algorithm}

Let $s_1,s_2,\ldots,s_e$ be chosen suitably and $r^e_{\max}$ be the maximum number of occurrence of each element of $V=\{1,2,\ldots,v\}$ in sub-blocks of $\cal B$ induced in step $e$, denoted by ${\cal B}_e$, and $k^{e}_{\max}$ be the maximum size of blocks of ${\cal B}_e$. Then, we must find all elements $0\le s\le m-1$ and all chains $i_0\to i_1\to i_2\to \ldots\to i_{l-1}\to i_{l}=i_0$, each $i_j,i_{j+1}$ is co-block in ${\cal B}_e$, such that all equations $\sum_{k=0}^{l}s_{i_k,i_{k+1}}$ ($1\le l\le \frac{g}{2}-1$, $s_{i_0,i_1}=s$) are not equal to zero in modulus of $m$. In step $e$ of the algorithm, two vertices $i_0$ and $i_1$ in the chains $i_0\to i_1\to i_2\to \ldots\to i_{l-1}\to i_{l}=i_0$ are given, so we must find all possible indices $i_2,\ldots,i_{l-1}$ to check whether the equations $\sum_{k=0}^{l}s_{i_k,i_{k+1}}$ are equal to zero or not. However, for each $2\le j\le l-1$, the number of choices $i_j$ is at most $(r^e_{\max}-1)(k^e_{\max}-1)$, because for each $1\le j\le l$, $i_j$ and $i_{j-1}$ is co-block and $i_j$ and $i_{j-2}$ is not co-block; hence if $i_{j-1}$ and $i_{j-2}$ are given from block $B$, then $i_j$ is one of co-block elements of $i_{j-1}$, except block $B$, whereas $i_{j-1}$ is repeated at most $r^e_{\max}-1$ in other blocks. Therefore, all possible vectors $(i_2,\ldots,i_{l-1})$ are enumerated at most $(r^e_{\max}-1)^{l-2}(k_{\max}-1)^{l-2}$, $l\le \frac{g}{2}-1$. But $0\le s\le m-1$. So, the complexity of step $e$ of the algorithm is $O((r^e_{\max}-1)^{\frac{g}{2}-3}(k^e_{\max}-1)^{\frac{g}{2}-3}m^e)$ and the overall complexity is about $\sum_{e=1}^{(k-1)b}O(r^e_{\max}-1)^{\frac{g}{2}-3}(k^e_{\max}-1)^{\frac{g}{2}-3}m^e)=O(b(k-1)^{\frac{g}{2}-2}(r-1)^{\frac{g}{2}-3})m^{b(k-1)})$, where $b=|{\cal B}|$, $r=\max r^{e}_{\max}$ and $k=\max k^{e}_{\max}$. For example, to find appropriate shift-sequence $S$ in Example 4 with $g({\cal H})=14$, the complexity is $O(10\times 2^5\times 4^4\times m^{20})=O(m^{20})$.

Therefore, the algorithm has a polynomial complexity with respect to the input value $m$, if $b$, $k$, $g$ and $r$ are fixed. On the other hand, if $g$ and $r$ are fixed, the complexity grows up exponentially with respect to $k$ or $b$. In this case, the exhaustive search is impractical and a heuristic method can be used to speed up the process of finding a satisfactory solution via selecting the shifts in each step randomly.
\section{Simulation Results} This section provides a bit-error rate
(BER) performance comparison over the AWGN channel
with BPSK modulation between some FSS codes with different
girths, on one hand, and random-like counterparts~\cite{mac}, PEG codes~\cite{peg}, cylinder-type LDPC codes~\cite{cylinder} and QC LDPC codes based on Steiner triple systems STS(9), STS(13) and the configuration Aff$^*$(16)~\cite{ghol4}  with the same rates and lengths, on the other
hand. The binary FSS codes have been decoded with
the iterative sum-product algorithm (SPA) and all randomly
generated codes given in this paper are constructed using Radford
Neal's software~\cite{Neal}, and they have as few number of 4-cycles as possible.
\medskip

In the performance figures, the notation FSS$(\lambda(x);gb)$, $c\in\{2,3\}$, refers to
a girth $b$-FSS code with column weight distribution $\lambda(x)$.
In addition, PEG$(n;gb)$ is used to denote the girth-$b$ LDPC code  of length $n$ constructed by PEG and QPEG$(x\times y;gb)$ is used to denote the girth-$b$ QC LDPC code obtained by applying the proposed algorithm in Section~\ref{secalg} to a $x\times y$ base matrix generated by PEG. Also, Rand$(n,cb)$ denotes a random-like column-weight $b$ code
of length $n$ constructed with Neal's software. Moreover, by C$(m;gb)$, STS13$(m;gb)$ and con12$(m;gb)$, we mean the cylinder-type QC LDPC codes \cite{cylinder} and QC LDPC codes based on STS(13) and Aff$^*$(16)~\cite{ghol4}, respectively. 
\medskip

\begin{figure}
\begin{center}
\includegraphics[scale=0.6]{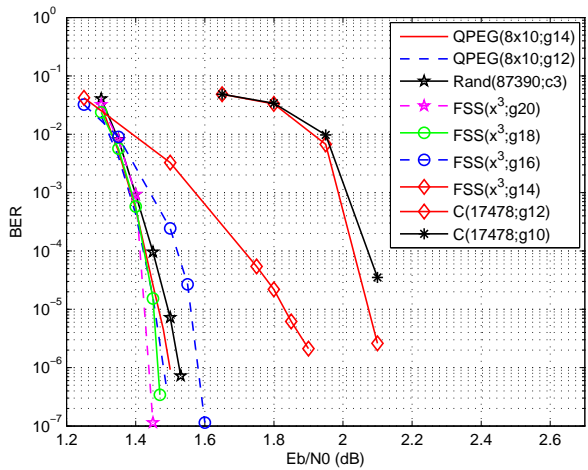}
\end{center}
\caption{\small A comparison between some FSS codes constructed by method I, with different
girths, random-like, QPEG and cylinder-type QC LDPC codes of length 87390 and rate about 0.2.
}\label{s1}
\end{figure}

\begin{figure}
\begin{center}
\includegraphics[scale=0.565]{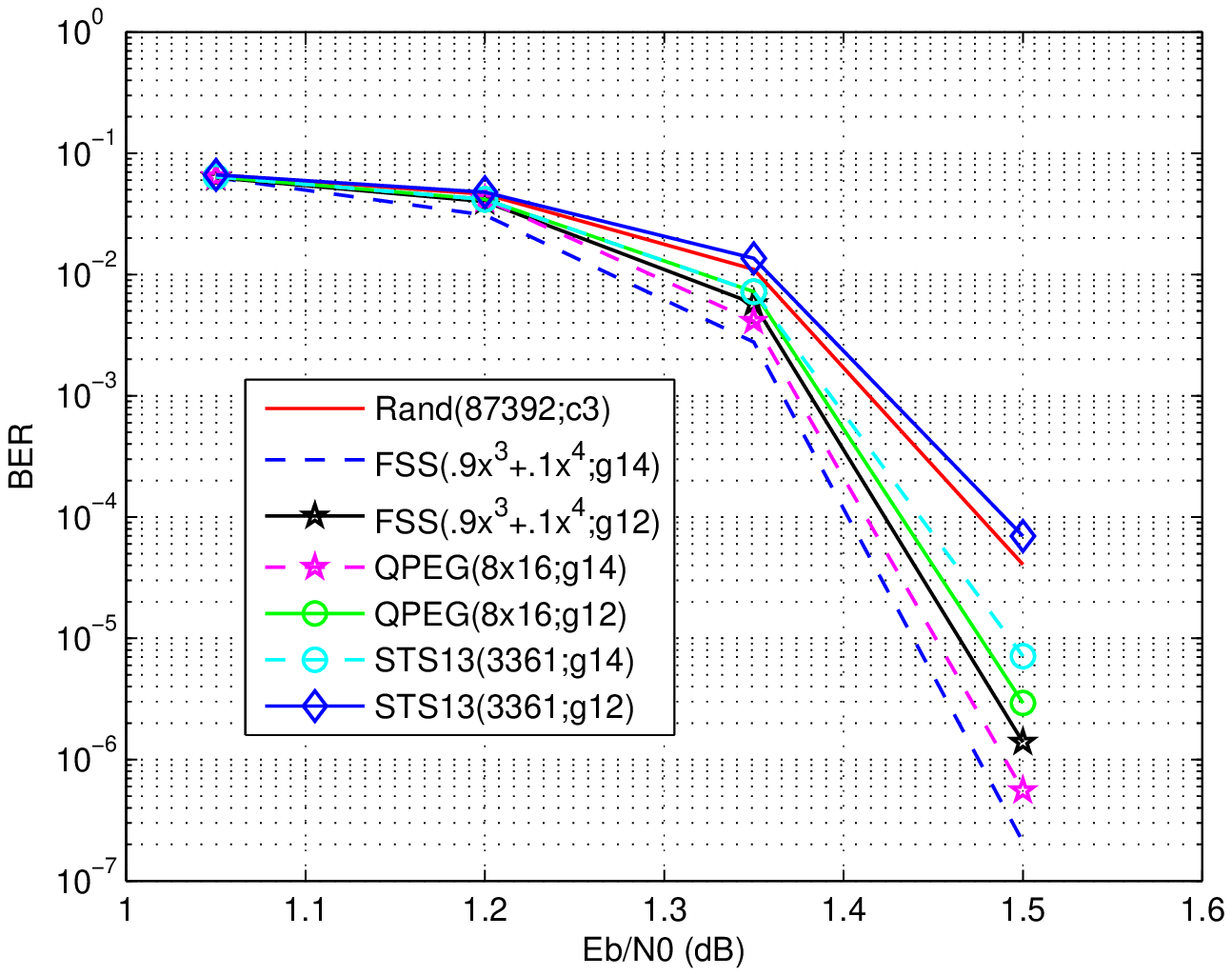}
\end{center}
\caption{\small  A comparison between some FSS codes constructed by method II, QPEG, STS(13) and random-like LDPC codes of length 87000 and rate about 0.5.
}\label{s2}
\end{figure}

\begin{figure}
\begin{center}
\includegraphics[scale=0.6]{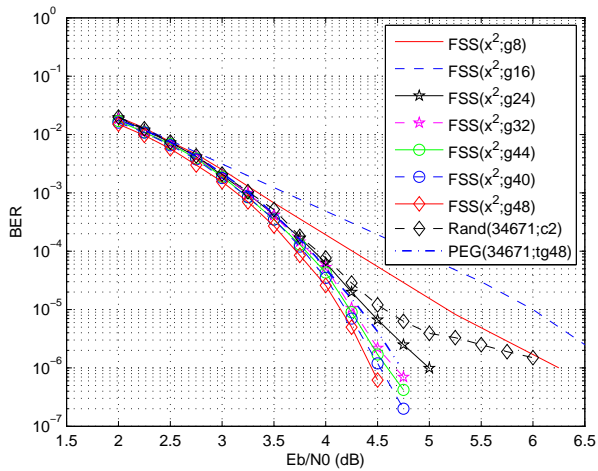}
\end{center}
\caption{\small A comparison between some column-weight two FSS codes with different girths,
random-like and PEG LDPC codes of length 34671 and rate about 0.26.} \label{c2}
\end{figure}

Although the high-length random codes usually have better performance rather than the structured ones,
Figure \ref{s1} shows that the FSS codes with girth 20 and 18 constructed by Method I outperform the QPEG and Random-like codes. Moreover, the constructed FSS codes have remarkably coding gains, about 0.6 dB, over than cylinder-type QC LDPC codes with girths 10 and 12.
Figure \ref{s2} shows that the FSS codes with girths 12 and 14 constructed by Method II outperform the QPEG codes with girth 12 and 14, respectively.
Also, the constructed FSS codes have coding gains over than random-like counterparts and QC LDPC codes with girths 12 and 14 based on STS(13).
Figure~\ref{c2} highlights the positive role of girth in improving the efficiency of performance of column-weight two FSS codes. Moreover, column-weight two FSS codes with girth at least 32 outperform PEG and random-like LDPC codes. Finally, Figure~\ref{c3,4} implies that the FSS codes with girths 14 and 16 outperform the random-like, QPEG, cylinder-type and Aff$^*$(16) based QC LDPC codes.


\medskip

The performance of an LDPC code depends not only on the girth of its Tanner graph, but also on the structure of the Tanner graph and the weight distribution of the parity-check matrix. Although PEG algorithm is one of the best powerful algorithm to generate LDPC codes with large girth and good weight distribution in short-block-length, simulation results confirm that the constructed codes with large girth outperform than PEG codes in medium lengths. Maybe it is because the real girth of the codes constructed by PEG algorithm is less than their target girth in medium lengths when the target girth is high. Moreover, in constructing FSS codes with large girth, some attempts have been made to construct base matrices with good weight distribution profiles. Therefore, FSS codes with large girth outperform than PEG and QPEG codes. Besides, there is considerable complexity to generate high girth PEG codes when the length of the code enlarges. PEG codes are not among the QC LDPC codes, so their encoding and decoding process has more complexity rather than FSS codes. QPEG is a good approach to construct QC LDPC codes having mother matrices constructed by PEG algorithm. Nevertheless, the weigh distribution of QPEG is not optimal as well as PEG codes.

\section{Conclusion}
In this paper, we proposed some binary 
well-structured LDPC codes based on some combinatorial designs,
called finite set systems. The girth and column-weight distribution of
parity-check matrices of these proposed codes are arbitrary. 
Also, for a given binary matrix $H$, we determined the maximum achievable girth of QC-LDPC codes
having mother matrix $H$. Then, we proposed two structured methods
to construct mother matrices such that their associated QC-LDPC codes
have large girths. Our constructed QC-LDPC codes with column-weight two have larger rates compared to the rate of
geometric and cylinder-type LDPC codes with the same girth.
Simulation results confirm that the constructed codes with larger girths perform better than the codes with smaller girths, and randomly constructed LDPC codes.
\section{ACKNOWLEDGMENT}
We would like to thank the anonymous referees for their
helpful comments. This work was supported in part by the research
council of Shahrekord university.


\begin{table}[h]
\caption{ Some FSS$(m, {\cal B}, S)$ codes with girth $2g$ and length $n$
derived from FSS$(v, b, 2)$}
\tiny
$\hspace{-1cm}
\begin{array}{|c|p{4cm}|c|c|c|c|p{8cm}|}
\hline \it  \it (v, b)&{\cal B}&{\cal R}&m&2g&n&S\\
\hline\hline
&&&3&8&90&[0, 0, 0, 0, 0, 0, 0, 0, 1, 0, 0, 0, 0, 0, 0, 0, 0, 0, 0, 0, 1, 0, 0, 1, 0, 2, 0, 0, 1, 0, 0, 1, 0, 1, 0, 0, 0, 2, 2, 1, 0, 1, 1, 2, 0]\\
\cline{4-7}
&&&8&10&240&[0, 0, 0, 0, 0, 0, 0, 0, 1, 0, 0, 0, 0, 0, 0, 0, 0, 0, 0, 0, 1, 1, 3, 4, 0, 2, 2, 0, 3, 1, 0, 1, 1, 5, 0, 2, 0, 4, 5, 3, 5, 3, 2, 7, 1]\\
\cline{4-7}
&&&10&12&300&[0, 0, 0, 0, 0, 0, 0, 0, 1, 0, 0, 0, 0, 0, 0, 0, 0, 0, 0, 0, 2, 2, 5, 6, 1, 2, 8, 0, 5, 3, 3, 7, 3, 9, 12, 2, 3, 10, 12, 5, 18, 18, 17, 23, 15]\\
\cline{4-7}
&[\{1, 2, 3\}, \{3, 11\}, \{2, 14\}, \{8, 17\}, \{10, 11\}, \{11, 16\}, \{10, 15, 16\},&&100&14&3000&[0, 0, 0, 0, 0, 0, 0, 0, 1, 0, 0, 2, 0, 0, 0, 0, 0, 0, 0, 0, 3, 2, 6, 10, 1, 2, 7, 0, 10, 2, 6, 18, 1, 16, 25, 3, 10, 27, 46, 5, 33, 41, 58, 81, 52]\\
\cline{4-7}
(18,30)&\{7, 17\}, \{7, 16\}, \{15, 17\}, \{12, 14, 18\}, \{5, 6, 9, 18\}, \{14, 17\}, \{4, 6\},&0.4&359&16&10770&[0, 0, 0, 0, 0, 0, 0, 0, 1, 0, 0, 3, 0, 0, 0, 0, 0, 1, 0, 1, 5, 7, 0, 13, 4, 32, 22, 0, 20, 16, 22, 47, 3, 62, 92, 12, 22, 53, 88, 7, 76, 113, 131, 198, 115]\\
\cline{4-7}
& \{6, 10, 14\}, \{4, 11, 17, 18\}, \{2, 4, 12\}, \{2, 10\}, \{4, 13\}, \{16, 17\}, \{8, 10\}, \{3, 9, 10\}, \{5, 8, 12, 16\}, \{7, 9, 13\}, \{4, 7\}, \{3, 7\}, \{1, 9, 15\},&&4000&18&120000&[0, 3447, 3447, 3447, 3447, 3447, 3447, 3447, 3447, 3447, 3447, 270, 3447, 3447, 3447, 3447, 3447, 270, 3447, 2793, 1875, 582, 228, 3573, 3224, 2570, 786, 1567, 1567, 1218, 2779, 446, 2417, 2750, 3743, 930, 2917, 3341, 3591, 1099, 3650, 1605, 1709, 2140, 1453]\\
\cline{4-7}
& \{8, 9\}, \{2, 7, 8\}, \{1, 18\}&&40000&20&1200000&[0, 8301, 19064, 19064, 19064, 19064, 19064, 19064, 19064, 19064, 19064, 5196, 19064, 19064, 19064, 19064, 19064, 19064, 29828, 29828, 26036, 24920, 8075, 11540, 15726, 27873, 28060, 15627, 15627, 11342, 8003, 22006, 18470, 15132, 1455, 7042, 9361, 7394, 11468, 4501, 20308, 8990, 1284, 26684, 15693] \\
\hline\hline
&&&4&8&108&[0, 0, 0, 0, 0, 0, 0, 0, 1, 1, 1, 0, 0, 0, 0, 1, 1, 2, 1, 2, 0, 0, 0, 1, 0, 1, 0, 2, 0, 2, 0, 0, 0]\\
\cline{4-7}
&&&7&10&189&[0, 5, 1, 1, 1, 1, 1, 1, 5, 5, 1, 1, 5, 2, 5, 2, 2, 2, 4, 1, 0, 4, 5, 0, 5, 0, 2, 1, 0, 3, 0, 0, 0]\\
\cline{4-7}
&[\{1, 2\}, \{2, 7\}, \{5, 13\}, \{1, 3\}, \{9, 14\}, \{2, 8, 12, 13\}, \{1, 13\}, \{9, 14\}, &&15&12&405&[0, 6, 4, 4, 4, 4, 4, 4, 11, 11, 4, 4, 4, 11, 12, 4, 4, 11, 5, 2, 11, 11, 2, 6, 2, 7, 6, 7, 7, 7, 3, 4, 3]\\
\cline{4-7}
(14,27)&\{7, 8\}, \{13, 14\}, \{4, 12\}, \{5, 6\}, \{5, 10\}, \{5, 10\}, \{6, 12\}, \{2, 3, 4, 6\}, &0.48&100&14&2700&[0, 16, 16, 16, 16, 16, 16, 16, 91, 91, 16, 16, 16, 16, 98, 32, 98, 98, 32, 53, 98, 35, 35, 71, 35, 35, 64, 74, 18, 60, 79, 94, 39] \\
\cline{4-7}
&\{4, 5\}, \{6, 8\}, \{1, 14\}, \{2, 10\}, \{10, 12\}, \{3, 7, 12, 14\}, \{4, 14\}, \{4, 8\}, &&175&16&4725&[0, 126, 126, 126, 126, 126, 126, 126, 161, 168, 20, 20, 20, 20, 20, 21, 20, 20, 21, 82, 142, 116, 158, 130, 150, 123, 31, 59, 87, 107, 93, 150, 118]\\
\cline{4-7}
&\{5, 11\}, \{3, 11\}, \{9, 10\}]&&700&18&18900&[0, 511, 397, 397, 397, 397, 397, 397, 104, 626, 684, 684, 570, 570, 570, 565, 570, 570, 321, 122, 258, 206, 394, 273, 219, 567, 506, 36, 122, 57, 415, 518, 23]\\
\hline\hline
& &&4&8&144&[0, 1, 1, 1, 1, 1, 1, 0, 1, 1, 1, 1, 0, 1, 1, 0, 0, 1, 1, 1, 0, 1, 1, 1, 0, 0, 1, 1, 1, 2, 0, 0, 1, 2, 0, 1, 2, 1, 0, 1, 2, 0, 1, 1, 0, 2, 1, 0]\\
\cline{4-7}
& &&13&10&468&[0, 8, 8, 8, 8, 8, 8, 6, 8, 8, 8, 8, 6, 8, 8, 6, 5, 8, 8, 8, 8, 8, 1, 6, 1, 0, 6, 1, 1, 6, 5, 7, 11, 4, 5, 0, 5, 4, 7, 8, 10, 2, 11, 10, 1, 11, 3, 0]\\
\cline{4-7}
&[\{1, 2\}, \{14, 15\}, \{6, 13, 14\}, \{8, 13\}, \{3, 6\}, \{2, 4\}, \{1, 4, 15\}, \{10, 12\}, \{5, 9\}, \{1, 9, 10\}, \{8, 14\}, &&40&12&1440&[0, 21, 14, 14, 14, 14, 14, 30, 14, 14, 14, 14, 30, 14, 14, 30, 14, 14, 14, 14, 14, 30, 5, 30, 5, 18, 18, 13, 18, 28, 22, 22, 13, 15, 23, 36, 37, 23, 19, 11, 32, 8, 14, 29, 3, 3, 13, 2]\\
\cline{4-7}
(15, 36)& \{11, 14\}, \{5, 7, 10\}, \{13, 15\}, \{4, 7, 8\}, \{5, 12, 14\}, \{5, 15\}, \{2, 12, 13\}, \{4, 13\}, \{11, 12\}, \{10, 13\}, \{6, 10, 11\}, \{2, 8\}, \{3, 7, 11\}, \{8, 15\}, &0.58&250&14&9000&[0, 126, 126, 126, 126, 126, 126, 161, 126, 126, 126, 126, 161, 126, 126, 161, 2, 126, 126, 215, 215, 215, 85, 85, 132, 55, 55, 109, 14, 104, 63, 152, 57, 201, 166, 28, 28, 221, 245, 199, 25, 184, 212, 119, 225, 118, 108, 15] \\
\cline{4-7}
& \{8, 11\}, \{1, 3, 14\}, \{3, 5\}, \{1, 12\}, \{4, 5, 6\}, \{2, 6\}, \{3, 9\}, \{8, 10\}, \{2, 9\}, \{9, 11, 15\}, \{1, 7\}]&&1000&16&36000&[0, 948, 785, 785, 785, 785, 785, 207, 785, 785, 785, 785, 207, 785, 785, 495, 785, 785, 785, 144, 901, 144, 144, 980, 980, 691, 534, 787, 88, 283, 465, 228, 945, 889, 125, 193, 242, 732, 538, 670, 790, 922, 478, 656, 901, 606, 720, 22]\\
\cline{4-7}
&&&2000&18&72000&[0, 12965, 12965, 12965, 12965, 12965, 9692, 11633, 9692, 9692, 9692, 9692, 9692, 9692, 9692, 16362, 16362, 9692, 9692, 16868, 5527, 16868, 773, 7948, 4676, 19659, 10739, 8995, 7882, 19593, 4544, 10607, 9494, 883, 3715, 5393, 15359, 13982, 3252, 9814, 7535, 12742, 9455, 7051, 17668, 18241, 8729, 11620]\\
\hline\hline
(3, 10)&[\{1, 2, 3\}, \{1, 2, 3\}, \{1, 2, 3\}, &0.7&36&8&360&[0, 0, 1, 2, 3, 6, 28, 35, 24, 33, 15, 30, 22, 14, 25, 13, 17, 21, 16, 11]\\
\cline{4-7}
&\{1, 2, 3\}, \{1, 2, 3\}, \{1, 2, 3\}, \{1, 2, 3\}, \{1, 2, 3\}, \{1, 2, 3\}, \{1, 2, 3\}]&&477&10&4770&[0, 0, 1, 3, 5, 13, 449, 466, 408, 446, 373, 427, 288, 369, 62, 343, 153, 320, 333, 125]\\
\cline{4-7}
&&&2570&12&25700&[0, 0, 1, 3, 7, 19, 2522, 2545, 2417, 2492, 2208, 2393, 2033, 2251, 293, 2128, 867, 1963, 992, 1696]\\
\hline\hline
&&&11&6&121&[0, 10, 10, 8, 9, 9, 8, 5, 7, 2, 6, 0, 5, 1, 4, 7, 3, 4, 2, 6, 1, 3]\\
\cline{4-7}
(3, 11)&[\{1, 2, 3\}, \{1, 2, 3\}, \{1, 2, 3\}, \{1, 2, 3\}, \{1, 2, 3\}, \{1, 2, 3\}, \{1, 2, 3\}, &0.73&44&8&484&[0, 43, 43, 41, 41, 37, 40, 35, 35, 24, 33, 21, 31, 13, 30, 5, 27, 3, 19, 4, 13, 0]\\
\cline{4-7}
&\{1, 2, 3\}, \{1, 2, 3\}, \{1, 2, 3\}, \{1, 2, 3\}]&&645&10&7095&[0, 0, 1, 3, 5, 13, 12, 29, 591, 626, 548, 606, 478, 557, 86, 533, 132, 519, 214, 396, 406, 248]\\
\cline{4-7}
&&&4000&12&44000&[0, 0, 1, 3, 7, 19, 28, 64, 75, 170, 101, 256, 296, 556, 376, 822, 505, 1168, 1004, 2305, 1276, 3260]\\
\hline\hline
&&&13&6&156&[0, 0, 1, 2, 2, 1, 3, 5, 4, 8, 5, 10, 7, 3, 8, 11, 6, 12, 9, 4, 10, 7, 11, 9]\\
\cline{4-7}
(3, 12)&[\{1, 2, 3\}, \{1, 2, 3\}, \{1, 2, 3\}, \{1, 2, 3\}, \{1, 2, 3\}, \{1, 2, 3\}, \{1, 2, 3\}, &0.75&51&8&612&[0, 0, 1, 2, 3, 6, 4, 11, 45, 49, 41, 50, 29, 43, 27, 40, 14, 39, 19, 37, 16, 35, 9, 24]\\
\cline{4-7}
&\{1, 2, 3\}, \{1, 2, 3\}, \{1, 2, 3\}, \{1, 2, 3\}, \{1, 2, 3\}, \{1, 2, 3\}]&&837&10&10044&[0, 0, 1, 3, 5, 13, 12, 29, 783, 818, 740, 798, 691, 773, 615, 711, 111, 683, 181, 611, 145, 522, 645, 417]\\
\cline{4-7}
&&&5100&12&61200&[0, 0, 1, 3, 7, 19, 28, 64, 4993, 5061, 4827, 4975, 4601, 4780, 4174, 4588, 426, 4518, 729, 3952, 2231, 3725, 2217, 3834]\\
\hline
\end{array}
\label{tab-code}$
\end{table}
\begin{table}[ht]
\caption{Some FSS$(v,b,2)$ obtained from a primitive FSS$(v', b', 2)$ constructed by method I.}
\scriptsize
\tiny
\[\hspace{-1.5cm}
\begin{array}{|c|c|p{3cm}|c|c||c|p{10cm}|c|}
\hline {\cal R}&(v', b')&{\cal B}' ({\rm Blocks~ of~ the~ primitive~ FSS})&2g'&g({\cal B}')&(v, b)&{\cal B}~ ({\rm Blocks~ of~ the~ new~ FSS})&g({\cal B})\\
\hline\hline
0.33&(2, 3)&[\{1, 2\}, \{1, 2\}, \{1, 2\}]&8&12&(6, 9)&[\{1, 4\}, \{2, 5\}, \{3, 6\}, \{1, 5\}, \{2, 6\}, \{3, 4\}, \{1, 6\}, \{2, 4\}, \{3, 5\}]&24\\
\hline
0.33&(2, 3)&[\{1, 2\}, \{1, 2\}, \{1, 2\}]&12&12&(14, 21)&[\{1, 8\}, \{2, 9\}, \{3, 10\}, \{4, 11\}, \{5, 12\}, \{6, 13\}, \{7, 14\}, \{1, 9\},
 \{2, 10\}, \{3, 11\}, \{4, 12\}, \{5, 13\}, \{6, 14\}, \{7, 8\}, \{1, 11\}, \{2, 12\},
\{3, 13\}, \{4, 14\}, \{5, 8\}, \{6, 9\}, \{7, 10\}]&36\\
\hline
0.5&(2, 4)&[\{1, 2\}, \{1, 2\}, \{1, 2\}, \{1, 2\}]&8&12&(8, 16)&[\{1, 5\}, \{2, 6\}, \{3, 7\}, \{4, 8\}, \{1, 6\}, \{2, 7\}, \{3, 8\}, \{4, 5\},
\{1, 7\}, \{2, 8\}, \{3, 5\}, \{4, 6\}, \{1, 8\}, \{2, 5\}, \{3, 6\}, \{4, 7\}]&24\\
\hline
0.5&(2, 4)&[\{1, 2\}, \{1, 2\}, \{1, 2\}, \{1, 2\}]&12&12&(26, 52)&[\{1, 14\}, \{2, 15\}, \{3, 16\}, \{4, 17\}, \{5, 18\}, \{6, 19\}, \{7, 20\}, \{8, 21\},\{9, 22\}, \{10, 23\}, \{11, 24\}, \{12, 25\}, \{13, 26\}, \{1, 15\}, \{2, 16\}, \{3, 17\},
 \{4, 18\}, \{5, 19\}, \{6, 20\}, \{7, 21\}, \{8, 22\}, \{9, 23\}, \{10, 24\}, \{11, 25\},
\{12, 26\}, \{13, 14\}, \{1, 17\}, \{2, 18\}, \{3, 19\}, \{4, 20\}, \{5, 21\}, \{6, 22\},
 \{7, 23\}, \{8, 24\}, \{9, 25\}, \{10, 26\}, \{11, 14\}, \{12, 15\}, \{13, 16\}, \{1, 23\},
 \{2, 24\}, \{3, 25\}, \{4, 26\}, \{5, 14\}, \{6, 15\}, \{7, 16\}, \{8, 17\}, \{9, 18\},
 \{10, 19\}, \{11, 20\}, \{12, 21\}, \{13, 22\}]&36\\
\hline
0.6&(2, 5)&[\{1, 2\}, \{1, 2\}, \{1, 2\}, \{1, 2\}, \{1, 2\}]&8&12&(10, 25)&[\{1, 6\}, \{2, 7\}, \{3, 8\}, \{4, 9\}, \{5, 10\}, \{1, 7\}, \{2, 8\}, \{3, 9\}, \{4, 10\}, \{5, 6\}, \{1, 8\}, \{2, 9\}, \{3, 10\}, \{4, 6\}, \{5, 7\}, \{1, 9\}, \{2, 10\}, \{3, 6\}, \{4, 7\}, \{5, 8\}, \{1, 10\}, \{2, 6\}, \{3, 7\}, \{4, 8\}, \{5, 9\}]&24\\
\hline
0.6&(2, 5)&[\{1, 2\}, \{1, 2\}, \{1, 2\}, \{1, 2\}, \{1, 2\}]&12&12&(46, 115)&[\{1, 24\}, \{2, 25\}, \{3, 26\}, \{4, 27\}, \{5, 28\}, \{6, 29\}, \{7, 30\}, \{8, 31\}, \{16, 39\}, \{17, 40\}, \{18, 41\}, \{19, 42\}, \{20, 43\}, \{21, 44\}, \{22, 45\},
 \{23, 46\}, \{1, 25\}, \{2, 26\}, \{3, 27\}, \{4, 28\}, \{5, 29\}, \{6, 30\}, \{7, 31\},
 \{8, 32\}, \{9, 33\}, \{10, 34\}, \{11, 35\}, \{12, 36\}, \{13, 37\}, \{14, 38\},
 \{15, 39\}, \{16, 40\}, \{17, 41\}, \{18, 42\}, \{19, 43\}, \{20, 44\}, \{21, 45\},
 \{22, 46\}, \{23, 24\}, \{1, 27\}, \{2, 28\}, \{3, 29\}, \{4, 30\}, \{5, 31\}, \{6, 32\},
 \{7, 33\}, \{8, 34\}, \{9, 35\}, \{10, 36\}, \{11, 37\}, \{12, 38\}, \{13, 39\}, \{14, 40\},
 \{15, 41\}, \{16, 42\}, \{17, 43\}, \{18, 44\}, \{19, 45\}, \{20, 46\}, \{21, 24\}, \{22, 25\},
\{23, 26\}, \{1, 32\}, \{2, 33\}, \{3, 34\}, \{4, 35\}, \{5, 36\}, \{6, 37\}, \{7, 38\},
 \{8, 39\}, \{9, 40\}, \{10, 41\}, \{11, 42\}, \{12, 43\}, \{13, 44\}, \{14, 45\}, \{15, 46\},
\{16, 24\}, \{17, 25\}, \{18, 26\}, \{19, 27\}, \{20, 28\}, \{21, 29\}, \{22, 30\}, \{23, 31\},
 \{1, 38\}, \{2, 39\}, \{3, 40\}, \{4, 41\}, \{5, 42\}, \{6, 43\}, \{7, 44\}, \{8, 45\},
\{9, 46\}, \{10, 24\}, \{11, 25\}, \{12, 26\}, \{13, 27\}, \{14, 28\}, \{15, 29\}, \{16, 30\},
\{17, 31\}, \{18, 32\}, \{19, 33\}, \{20, 34\}, \{21, 35\}, \{22, 36\}, \{23, 37\}]&36\\
\hline
0.66&(2, 6)&[\{1, 2\}, \{1, 2\}, \{1, 2\}, \{1, 2\}, \{1, 2\}, \{1, 2\}]&8&12&(12, 36)&[\{1, 7\}, \{2, 8\}, \{3, 9\}, \{4, 10\}, \{5, 11\}, \{6, 12\}, \{1, 8\}, \{2, 9\}, \{3, 10\}, \{4, 11\}, \{5, 12\}, \{6, 7\}, \{1, 9\}, \{2, 10\}, \{3, 11\}, \{4, 12\},
\{5, 7\}, \{6, 8\}, \{1, 10\}, \{2, 11\}, \{3, 12\}, \{4, 7\}, \{5, 8\}, \{6, 9\},
\{1, 11\}, \{2, 12\}, \{3, 7\}, \{4, 8\}, \{5, 9\}, \{6, 10\}, \{1, 12\}, \{2, 7\},
 \{3, 8\}, \{4, 9\}, \{5, 10\}, \{6, 11\}]&24\\
\hline
0.25&(3, 4)&[\{1, 2, 3\}, \{1, 2, 3\}, \{1, 2, 3\}, \{1, 2, 3\}]&6&12&(15, 20)&[\{1, 6, 11\}, \{2, 7, 12\}, \{3, 8, 13\}, \{4, 9, 14\}, \{5, 10, 15\}, \{1, 7, 13\}, \{2, 8,  14\}, \{3, 9, 15\}, \{4, 10, 11\}, \{5, 6, 12\}, \{1, 8, 15\}, \{2, 9, 11\}, \{3, 10, 12\},\{4, 6, 13\}, \{5, 7, 14\}, \{1, 9, 12\}, \{2, 10, 13\}, \{3, 6, 14\}, \{4, 7, 15\}, \{5, 8, 11\}]&18\\
\hline
0.4&(3, 5)&[\{1, 2, 3\}, \{1, 2, 3\}, \{1, 2, 3\}, \{1, 2, 3\}, \{1, 2, 3\}]&6&12&(15, 25)&[\{1, 6, 11\}, \{2, 7, 12\}, \{3, 8, 13\}, \{4, 9, 14\}, \{5, 10, 15\}, \{1, 7, 13\}, \{2, 8,  14\}, \{3, 9, 15\}, \{4, 10, 11\}, \{5, 6, 12\}, \{1, 8, 15\}, \{2, 9, 11\}, \{3, 10, 12\}, \{4, 6, 13\},
\{5, 7, 14\}, \{1, 9, 12\}, \{2, 10, 13\}, \{3, 6, 14\}, \{4, 7, 15\}, \{5, 8, 11\}, \{1, 10, 14\},
\{2, 6, 15\}, \{3, 7, 11\}, \{4, 8, 12\}, \{5, 9, 13\}]&18\\
\hline
0.4&(3, 5)&[\{1, 2, 3\}, \{1, 2, 3\}, \{1, 2, 3\}, \{1, 2, 3\}, \{1, 2, 3\}]&8&12&(42, 70)&[\{1, 15, 29\}, \{2, 16, 30\}, \{3, 17, 31\}, \{4, 18, 32\}, \{5, 19, 33\}, \{6, 20, 34\},  \{7, 21, 35\},
 \{8, 22, 36\}, \{9, 23, 37\}, \{10, 24, 38\}, \{11, 25, 39\}, \{12, 26, 40\}, \{13, 27, 41\}, \{14, 28, 42\},
\{1, 16, 31\}, \{2, 17, 32\}, \{3, 18, 33\}, \{4, 19, 34\}, \{5, 20, 35\}, \{6, 21, 36\}, \{7, 22, 37\},
 \{8, 23, 38\}, \{9, 24, 39\}, \{10, 25, 40\}, \{11, 26, 41\}, \{12, 27, 42\}, \{13, 28, 29\}, \{14, 15, 30\},
\{1, 18, 36\}, \{2, 19, 37\}, \{3, 20, 38\}, \{4, 21, 39\}, \{5, 22, 40\}, \{6, 23, 41\}, \{7, 24, 42\},
\{8, 25, 29\}, \{9, 26, 30\}, \{10, 27, 31\}, \{11, 28, 32\}, \{12, 15, 33\}, \{13, 16, 34\}, \{14, 17, 35\},
 \{1, 19, 41\}, \{2, 20, 42\}, \{3, 21, 29\}, \{4, 22, 30\}, \{5, 23, 31\}, \{6, 24, 32\}, \{7, 25, 33\},
\{8, 26, 34\}, \{9, 27, 35\}, \{10, 28, 36\}, \{11, 15, 37\}, \{12, 16, 38\}, \{13, 17, 39\}, \{14, 18, 40\},
  \{1, 20, 39\}, \{2, 21, 40\}, \{3, 22, 41\}, \{4, 23, 42\}, \{5, 24, 29\}, \{6, 25, 30\}, \{7, 26, 31\},
\{8, 27, 32\}, \{9, 28, 33\}, \{10, 15, 34\}, \{11, 16, 35\}, \{12, 17, 36\}, \{13, 18, 37\}, \{14, 19, 38\}]&24\\
\hline
0.4&(3, 6)&[\{1, 2, 3\}, \{1, 2, 3\}, \{1, 2, 3\}, \{1, 2, 3\}, \{1, 2, 3\}, \{1, 2, 3\}]&6&12&(21, 42)&[\{1, 8, 15\}, \{2, 9, 16\}, \{3, 10, 17\}, \{4, 11, 18\}, \{5, 12, 19\}, \{6, 13, 20\}, \{7,  14, 21\},
 \{1, 9, 17\}, \{2, 10, 18\}, \{3, 11, 19\}, \{4, 12, 20\}, \{5, 13, 21\}, \{6, 14, 15\}, \{7, 8, 16\},
  \{1, 10, 19\}, \{2, 11, 20\}, \{3, 12, 21\}, \{4, 13, 15\}, \{5, 14, 16\}, \{6, 8, 17\}, \{7, 9, 18\},
  \{1, 11, 21\}, \{2, 12, 15\}, \{3, 13, 16\}, \{4, 14, 17\}, \{5, 8, 18\}, \{6, 9, 19\}, \{7, 10, 20\},
  \{1, 12, 16\}, \{2, 13, 17\}, \{3, 14, 18\}, \{4, 8, 19\}, \{5, 9, 20\}, \{6, 10, 21\}, \{7, 11, 15\},
 \{1, 13, 18\}, \{2, 14, 19\}, \{3, 8, 20\}, \{4, 9, 21\}, \{5, 10, 15\}, \{6, 11, 16\}, \{7, 12, 17\}]&18\\
\hline
0.33&(4, 6)&[\{1, 2, 3, 4\}, \{1, 2, 3, 4\}, \{1, 2, 3, 4\}, \{1, 2, 3, 4\},\{1, 2, 3, 4\}, \{1, 2, 3, 4\}] &6&12&(28, 42)&[\{1, 8, 15, 22\}, \{2, 9, 16, 23\}, \{3, 10, 17, 24\}, \{4, 11, 18, 25\}, \{5, 12, 19, 26\},  \{6, 13, 20, 27\},
 \{7, 14, 21, 28\}, \{1, 9, 17, 25\}, \{2, 10, 18, 26\}, \{3, 11, 19, 27\}, \{4, 12, 20, 28\}, \{5, 13, 21, 22\}, ,   \{6, 14, 15, 23\}, \{7, 8, 16, 24\}, \{1, 10, 19, 28\}, \{2, 11, 20, 22\}, \{3, 12, 21, 23\}, \{4, 13, 15, 24\},
  \{5, 14, 16, 25\}, \{6, 8, 17, 26\}, \{7, 9, 18, 27\}, \{1, 11, 21, 24\}, \{2, 12, 15, 25\}, \{3, 13, 16, 26\},
  \{4, 14, 17, 27\}, \{5, 8, 18, 28\}, \{6, 9, 19, 22\}, \{7, 10, 20, 23\}, \{1, 12, 16, 27\}, \{2, 13, 17, 28\},
 \{3, 14, 18, 22\}, \{4, 8, 19, 23\}, \{5, 9, 20, 24\}, \{6, 10, 21, 25\}, \{7, 11, 15, 26\}, \{1, 13, 18, 23\},
 \{3, 14, 18, 22\}, \{4, 8, 19, 23\}, \{5, 9, 20, 24\}, \{6, 10, 21, 25\}, \{7, 11, 15, 26\}, \{1, 13, 18, 23\},
 \{2, 14, 19, 24\}, \{3, 8, 20, 25\}, \{4, 9, 21, 26\}, \{5, 10, 15, 27\}, \{6, 11, 16, 28\}, \{7, 12, 17, 22\}]&18\\
\hline
\end{array}\]
\label{tab-struc1}
\end{table}
\begin{table}[ht]
\caption{Some new FSS$(v,b,2)$ obtained from a primitive FSS$(v',b',2)$ constructed by method II.}
\tiny
\[\hspace{-1.5cm}
\begin{array}{|c|c|c|p{3.5cm}|p{12.5cm}|}
\hline
(v,b)&g({\cal B})&{\cal R}&column-weight&{\cal B}\\
\hline\hline
(10,19)&14&0.47&[3, 3, 3, 4, 3, 3, 3, 4, 4, 3, 3, 3, 3, 4, 3, 3, 4, 3, 3]&[\{2, 3, 7\}, \{3, 4, 10\}, \{3, 4, 9\}, \{1, 3, 6, 10\}, \{1, 5, 6\}, \{1, 2, 4\}, \{4, 7, 10\}, \{2, 6, 9, 10\},  \{2, 5, 6, 8\}, \{2, 8, 10\},\{1, 7, 8\}, \{1, 4, 9\}, \{5, 7, 10\}, \{3, 6, 8, 9\}, \{1, 7, 9\}, \{2, 7, 9\}, \{5, 8, 9, 10\}, \{4, 5, 7\}, \{1, 3, 8\}]\\
\hline
(10, 23)&14&0.57&[2, 4, 2, 2, 4, 2, 2, 4, 2, 2, 4, 2, 4, 2, 4, 2, 4, 2, 2, 2, 2, 4, 2]&[\{1, 2\}, \{1, 5, 7, 10\}, \{1, 7\}, \{3, 10\}, \{3, 5, 9, 10\}, \{4, 8\}, \{2, 10\}, \{2, 4, 9, 10\}, \{6, 9\}, \{7, 9\}, \{2, 6, 8, 9\}, \{1, 2\}, \{1, 4, 8, 10\}, \{3, 6\}, \{3, 4, 7, 9\}, \{2, 3\}, \{4, 5, 6, 7\}, \{4, 5\}, \{1, 3\}, \{4, 6\}, \{6, 10\}, \{2, 3, 6, 7\}, \{1, 9\}]\\
\hline
(10, 24)&14&0.59&[3, 2, 2, 2, 2, 2, 2, 4, 3, 3, 4, 2, 2, 3, 4, 3, 2, 2, 4, 3, 3, 3, 2, 2]&[\{1, 2, 3\}, \{6, 7\}, \{4, 8\}, \{8, 9\}, \{8, 9\}, \{5, 8\}, \{6, 7\}, \{2, 7, 9, 10\}, \{4, 7, 8\}, \{6, 9, 10\}, \{1, 4, 7, 9\}, \{5, 9\}, \{3, 10\}, \{2, 6, 10\}, \{2, 3, 4, 5\}, \{1, 3, 4\}, \{2, 9\}, \{1, 2\}, \{2, 5, 7, 8\}, \{4, 5, 6\}, \{3, 8, 10\}, \{5, 7, 10\}, \{2, 6\}, \{3, 6\}]\\
\hline
(15, 40)&14&0.62&[2, 3, 2, 3, 3, 2, 2, 3, 3, 2, 3, 2 ,3, 2, 3, 3, 2, 3, 3, 2, 2, 3, 3, 3, 2, 3, 2 ,3, 2, 3, 3, 2, 2, 3, 2, 2, 2, 3, 3, 2]&[\{1, 2\}, \{5, 7, 13\}, \{7, 8\}, \{8, 10, 11\}, \{9, 12, 14\}, \{5, 11\}, \{12, 14\}, \{4, 8, 9\}, \{1, 10, 11\}, \{9, 12\}, \{1, 2, 7\}, \{1, 13\}, \{8, 9, 14\}, \{8, 15\}, \{3, 10, 14\}, \{11, 12, 13\}, \{2, 15\},  \{1, 8, 11\}, \{5, 12, 15\}, \{3, 9\},\{1, 3\}, \{3, 5, 7\}, \{5, 8, 13\}, \{4, 7, 15\}, \{13, 15\}, \{3, 14, 15\}, \{10, 14\}, \{7, 10, 13\},   \{13, 14\},  \{5, 10, 15\},\{9, 11, 13\}, \{4, 5\}, \{14, 15\}, \{2, 4, 5\}, \{9, 11\}, \{2, 7\}, \{10, 15\}, \{3, 4, 10\}, \{2, 3, 6\}, \{11, 15\}]\\
\hline
\hline
(10,13)&16&0.23&[3, 3, 3, 3, 3, 3, 4, 3, 3, 3, 3, 3, 3]&[\{1, 2, 3\}, \{2, 7, 8\}, \{7, 8, 10\}, \{1, 4, 8\}, \{3, 5, 8\}, \{6, 7, 9\}, \{1, 4, 9, 10\}, \{3, 9, 10\}, \{5, 8, 9\},  \{2, 6, 9\}, \{3, 4, 6\}, \{3, 4, 7\}, \{1, 5, 6\}]\\
\hline
(10, 21)&16&0.52&[2, 2, 2, 2, 3, 2, 4, 3, 2, 2, 2, 3, 2, 2, 4, 3, 2, 2, 4, 3, 3]&[\{3, 4\}, \{5, 10\}, \{4, 9\}, \{8, 9\}, \{3, 5, 6\}, \{6, 9\}, \{2, 4, 9, 10\}, \{5, 7, 10\}, \{3, 8\},  \{2, 8\}, \{7, 9\}, \{5, 8, 9\}, \{7, 8\}, \{2, 6\}, \{1, 4, 5, 6\}, \{2, 6, 7\}, \{7, 8\}, \{4, 8\}, \{1, 3, 7, 9\}, \{3, 8, 10\}, \{1, 2, 8\}]\\
\hline
(15, 40)&16&0.62&[2, 3, 2, 3, 3, 2, 2, 3, 3, 2, 3, 2, 3, 2, 3, 3, 2, 3, 3, 2, 2, 3, 3, 3, 2, 3, 2 ,3 ,2 , 3, 3, 2, 2, 3, 2, 2, 2, 3, 3, 2]&[\{1, 2\}, \{1, 4, 7\}, \{5, 15\}, \{10, 12, 13\}, \{2, 5, 10\}, \{2, 11\}, \{5, 14\}, \{7, 11, 14\}, \{10, 13, 14\}, \{9, 11\}, \{9, 11, 15\}, \{7, 15\}, \{2, 9, 10\}, \{7, 12\}, \{6, 8, 9\}, \{4, 5, 9\}, \{2, 11\}, \{3, 5, 14\}, \{4, 8, 10\}, \{2, 13\}, \{5, 8\}, \{5, 6, 12\}, \{8, 11, 13\}, \{6, 10, 15\}, \{8, 12\}, \{3, 8, 11\}, \{1, 2\}, \{1, 5, 7\}, \{6, 9\}, \{1, 8, 12\}, \{2, 6, 14\}, \{6, 14\}, \{8, 15\}, \{1, 14, 15\}, \{5, 8\}, \{2, 7\}, \{9, 14\}, \{4, 12, 14\}, \{3, 12, 15\}, \{7, 8\}]\\
\hline
(14, 39)&16&0.64&[2, 2, 2, 2, 2, 4, 2, 2, 2, 2, 2, 2, 2, 2, 2, 4, 2, 2, 2, 2, 2, 4, 2, 2, 2, 2, 2, 4, 2, 2, 2, 2, 4, 2, 2, 2, 2, 2, 2]&[\{1, 2\}, \{9, 13\}, \{12, 13\}, \{3, 13\}, \{9, 11\}, \{4, 11, 12, 13\}, \{9, 10\}, \{1, 11\}, \{2, 13\}, \{4, 11\}, \{13, 14\}, \{8, 9\}, \{6, 10\}, \{1, 8\}, \{2, 11\}, \{4, 5, 10, 14\}, \{3, 7\}, \{6, 10\}, \{4, 14\}, \{8, 10\}, \{7, 9\}, \{6, 7, 13, 14\}, \{8, 10\}, \{2, 13\}, \{8, 12\}, \{8, 11\}, \{7, 8\}, \{1, 3, 5, 11\}, \{1, 6\}, \{2, 6\}, \{6, 9\}, \{1, 12\}, \{1, 4, 7, 9\}, \{8, 14\}, \{9, 10\}, \{2, 11\}, \{5, 8\}, \{2, 7\}, \{3, 12\}]\\
\hline
\hline
(12, 24)&18&0.5&[3, 2, 2, 2, 2, 2, 2, 4, 3, 2, 2, 2, 3, 4, 2, 2, 3, 4, 3, 2, 2, 2, 2, 3]&[\{1, 2, 3\}, \{7, 9\}, \{6, 8\}, \{11, 12\}, \{4, 10\}, \{5, 8\}, \{1, 9\}, \{3, 4, 12\}, \{9, 11\}, \{2, 10\}, \{10, 12\}, \{1, 9, 10\}, \{3, 6, 8, 11\}, \{10, 11\}, \{4, 11\}, \{7, 11\}, \{8, 9, 12\}, \{2, 5, 7, 12\}, \{2, 4, 8\}, \{4, 5\}, \{5, 9\}, \{5, 11\}, \{1, 4\}, \{3, 7, 10\}]\\
\hline
(15, 30)&18&0.5&[2, 3, 2, 3, 3, 2, 2, 3, 3, 2, 3, 2, 3, 2, 3, 3, 2, 3, 3, 2, 2, 3, 3, 3, 2, 3, 2, 3, 2, 3]&[\{1, 2\}, \{8, 9, 15\}, \{12, 14\}, \{1, 13, 15\}, \{10, 15\}, \{10, 13\}, \{5, 10\}, \{3, 7, 11\}, \{13, 14\}, \{14, 15\}, \{5, 12, 15\}, \{1, 7\}, \{10, 11\}, \{1, 4, 15\}, \{3, 8\}, \{5, 9, 13\}, \{2, 10\}, \{2, 6, 10\}, \{11, 14\}, \{9, 10, 14\}, \{2, 12\}, \{6, 12\}, \{6, 8, 9\}, \{3, 13\}, \{4, 9, 12\}, \{11, 13\}, \{6, 13\}, \{7, 9, 11\}, \{12, 13\}, \{4, 7\}]\\
\hline
(14, 27)&18&0.48&[2, 2, 2, 2, 2, 4, 2, 2, 2, 2, 2, 2, 2, 2, 2, 4, 2, 2, 2, 2, 2, 4, 2, 2, 2, 2, 2]&[\{1, 2\}, \{2, 7\}, \{5, 13\}, \{1, 3\}, \{9, 14\}, \{2, 8, 12, 13\}, \{1, 13\}, \{9, 14\},\{7, 8\},\{13, 14\},  \{4, 12\}, \{5, 6\}, \{5, 10\}, \{5, 10\}, \{6, 12\}, \{2, 3, 4, 6\}, \{4, 5\}, \{6, 8\}, \{1, 14\}, \{2, 10\}, \{10, 12\}, \{3, 7, 12, 14\}, \{4, 14\}, \{4, 8\}, \{5, 11\}, \{3, 11\}, \{9, 10\}]\\
\hline
(20, 29)&18&0.31&[3, 3, 3, 3, 3, 3, 4, 3, 3, 3, 3, 3, 3, 3, 3, 3, 3, 3, 3, 3, 3, 3, 3, 4, 3, 3, 3, 3, 3]&[\{1, 2, 3\}, \{5, 13, 18\}, \{2, 4, 7\}, \{11, 13, 16\}, \{2, 18, 20\}, \{2, 4, 16\}, \{1, 9, 18, 19\}, \{15, 19, 20\}, \{9, 10, 20\}, \{2, 9, 11\}, \{12, 15, 19\}, \{4, 5, 17\}, \{13, 16, 17\}, \{8, 13, 14\}, \{7, 11, 20\}, \{3, 16, 19\}, \{2, 8, 14\}, \{10, 17, 20\}, \{5, 14, 20\}, \{7, 17, 18\}, \{3, 7, 8\}, \{6, 8, 18\}, \{4, 6, 15\}, \{1, 6, 10, 16\}, \{1, 5, 7\}, \{3, 12, 17\}, \{1, 11, 12\}, \{8, 17, 19\}, \{5, 6, 9\}]\\
\hline\hline
(20,25)&20&0.2&[2, 4, 2, 2, 4, 2, 2, 4, 2, 2, 4, 2, 4, 2, 4, 2, 4, 2, 2, 2, 2, 4, 2, 4, 2]&[ \{1, 2\}, \{4, 10\}, \{11, 14, 16, 18\}, \{14, 18\}, \{1, 9, 17, 20\}, \{6, 19\}, \{7, 9, 10, 16\}, \{6, 10\}, \{4, 10\}, \{14, 20\}, \{3, 7\}, \{3, 13\}, \{10, 12, 19, 20\}, \{4, 17\}, \{8, 15\}, \{10, 13\}, \{5, 11, 12, 17\}, \{2, 11, 13, 15\}, \{5, 19\}, \{1, 4, 8, 18\}, \{7, 19\}, \{6, 7, 15, 17\}, \{3, 18\}, \{5, 8, 9, 13\}, \{6, 8\} ]\\
\hline
(18,30)&20&0.4&[3, 2, 2, 2, 2, 2, 3, 2, 2, 2, 3, 4, 2, 2, 3, 4, 3, 2, 2, 2, 2, 3, 4, 3, 2, 2, 3, 2, 3, 2]&[\{1, 2, 3\}, \{3, 11\}, \{2, 14\}, \{8, 17\}, \{10, 11\}, \{11, 16\}, \{10, 15, 16\}, \{7, 17\}, \{7, 16\}, \{15, 17\}, \{12, 14, 18\}, \{5, 6, 9, 18\}, \{14, 17\}, \{4, 6\}, \{6, 10, 14\}, \{4, 11, 17, 18\}, \{2, 4, 12\}, \{2, 10\}, \{4, 13\}, \{16, 17\}, \{8, 10\}, \{3, 9, 10\}, \{5, 8, 12, 16\}, \{7, 9, 13\}, \{4, 7\}, \{3, 7\}, \{1, 9, 15\}, \{8, 9\}, \{2, 7, 8\},\{1,18\} ]\\
\hline
(15,36)&20&0.58&[2, 2, 3, 2, 2, 2, 3, 2, 3, 2, 2, 3, 2, 3, 2, 3, 2, 3, 2, 2, 2, 3, 2, 3, 2, 2, 3, 2, 2, 3]&[\{1, 2\}, \{14, 15\}, \{6, 13, 14\}, \{8, 13\}, \{3, 6\}, \{2, 4\}, \{1, 4, 15\}, \{5, 9\}, \{1, 9, 10\}, \{8, 14\}, \{11, 14\}, \{5, 7, 10\}, \{13, 15\}, \{4, 7, 8\}, \{10, 12\}, \{5, 12, 14\}, \{5, 15\}, \{2, 12, 13\}, \{4, 13\}, \{11, 12\}, \{10, 13\}, \{6, 10, 11\}, \{2, 8\}, \{3, 7, 11\}, \{8, 15\}, \{8, 11\}, \{1, 3, 14\}, \{3, 5\}, \{1, 12\}, \{4, 5, 6\}, \{2, 6\}, \{3, 9\}, \{8, 10\}, \{2, 9\}, \{9, 11, 15\}, \{1, 7\}]\\
\hline
\end{array}\]
\label{tab-struc2-1}
\end{table}
\begin{table}[ht]
\caption{New column-weight two FSS$(v,b,2)$ of rate $\cal R$ against the best previous found rate $r$.}
\scriptsize
\tiny
\[\hspace{-1.2cm}
\begin{array}{|c|c|c|c|p{15cm}|}
\hline
(v,b)&g({\cal B})&{\cal R}&r&${\cal B}$\\
\hline\hline
(7,11)&24&0.36&&[\{1, 2\}, \{4, 6\}, \{1, 6\}, \{2, 7\}, \{4, 7\}, \{1, 3\}, \{4, 7\}, \{5, 7\}, \{3, 5\}, \{1, 6\}, \{3, 5\}]\\
\cline{1-3}\cline{5-5}
(8,13)&24&0.38&&[\{1, 2\}, \{2, 3\}, \{2, 8\}, \{6, 7\}, \{1, 4\}, \{1, 3\}, \{5, 8\}, \{7, 8\}, \{5, 6\}, \{1, 6\}, \{4, 7\}, \{4, 8\}, \{2, 5\}]\\\cline{1-3}\cline{5-5}
(9,16)&24&0.44&0.33&[\{1, 2\}, \{3, 6\}, \{3, 6\}, \{8, 9\}, \{5, 7\}, \{7, 9\}, \{5, 6\}, \{7, 9\}, \{4, 7\}, \{1, 9\}, \{2, 8\}, \{3, 8\}, \{1, 6\}, \{2, 4\}, \{2, 5\}, \{4, 6\}]\\
\cline{1-3}\cline{5-5}
(12,26)&24&0.54&\cite{ghol2}&[\{1, 2\}, \{8, 12\}, \{3, 8\}, \{3, 4\}, \{9, 10\}, \{6, 12\}, \{7, 11\}, \{2, 12\}, \{8, 9\}, \{6, 12\}, \{3, 8\}, \{10, 11\}, \{11, 12\}, \{4, 11\}, \{2, 7\}, \{7, 8\}, \{9, 11\}, \{1, 10\}, \{5, 6\}, \{5, 7\}, \{2, 9\}, \{4, 5\}, \{2, 4\}, \{1, 5\}, \{5, 10\}, \{1, 8\}]
\\\cline{1-3}\cline{5-5}
(14,33)&24&0.58&&[\{1, 2\}, \{1, 9\}, \{9, 11\}, \{13, 14\}, \{12, 13\}, \{1, 14\}, \{12, 13\}, \{5, 11\}, \{6, 12\}, \{8, 14\}, \{4, 9\}, \{4, 5\}, \{11, 14\}, \{4, 6\}, \{10, 12\}, \{10, 11\}, \{9, 13\}, \{5, 6\}, \{7, 14\}, \{2, 13\}, \{6, 8\}, \{2, 5\}, \{1, 6\}, \{4, 10\}, \{3, 7\}, \{1, 10\}, \{8, 11\}, \{5, 7\}, \{3, 11\}, \{3, 9\}, \{7, 10\}, \{3, 6\}, \{2, 8\}]
\\
\hline\hline
(14,21)&32&0.33&&[\{1, 2\}, \{2, 10\}, \{12, 13\}, \{11, 14\}, \{6, 9\}, \{6, 12\}, \{4, 12\}, \{5, 10\}, \{3, 14\}, \{6, 8\}, \{4, 14\}, \{8, 12\}, \{1, 11\}, \{10, 13\}, \{1, 7\}, \{5, 9\}, \{7, 8\}, \{2, 4\}, \{9, 11\}, \{3, 5\}, \{3, 7\}]
\\
\cline{1-3}\cline{5-5}
(15,25)&32&0.4&&[\{3, 12\}, \{12, 13\}, \{14, 15\}, \{4, 12\}, \{5, 7\}, \{1, 3\}, \{4, 7\}, \{13, 14\}, \{9, 10\}, \{10, 15\}, \{6, 14\}, \{2, 8\}, \{10, 11\}, \{1, 2\}, \{4, 6\}, \{5, 15\}, \{5, 13\}, \{1, 14\}, \{5, 8\}, \{3, 9\}, \{6, 9\}, \{7, 11\}, \{2, 4\}, \{3, 11\}, \{8, 9\}]
\\
\cline{1-3}\cline{5-5}
(16,27)&32&0.41&0.25&[\{3, 13\}, \{4, 11\}, \{9, 11\}, \{9, 12\}, \{2, 11\}, \{6, 14\}, \{6, 9\}, \{4, 16\}, \{7, 14\}, \{3, 10\}, \{4, 5\}, \{5, 15\}, \{13, 16\}, \{8, 12\}, \{9, 10\}, \{12, 13\}, \{1, 14\}, \{8, 11\}, \{10, 15\}, \{1, 12\}, \{6, 16\}, \{7, 8\}, \{2, 3\}, \{2, 14\}, \{1, 15\}, \{5, 7\}, \{15, 16\}]
\\
\cline{1-3}\cline{5-5}
(17,30)&32&0.43&\cite{cylinder}&[\{10, 15\}, \{16, 17\}, \{4, 5\}, \{2, 7\}, \{8, 11\}, \{6, 8\}, \{1, 4\}, \{10, 13\}, \{8, 9\}, \{5, 16\}, \{1, 2\}, \{4, 14\}, \{5, 13\}, \{7, 9\}, \{1, 15\}, \{6, 17\}, \{10, 14\}, \{11, 16\}, \{1, 6\}, \{2, 13\}, \{11, 14\}, \{7, 14\}, \{3, 4\}, \{15, 16\}, \{12, 13\}, \{11, 12\}, \{7, 17\}, \{5, 9\}, \{3, 12\}, \{3, 6\}]
\\
\cline{1-3}\cline{5-5}
(19,34)&32&0.44&&[\{4, 10\}, \{18, 19\}, \{2, 5\}, \{4, 11\}, \{9, 19\}, \{2, 16\}, \{4, 6\}, \{7, 15\}, \{14, 16\}, \{6, 17\}, \{8, 19\}, \{3, 7\}, \{1, 3\}, \{15, 16\}, \{5, 10\}, \{1, 11\}, \{11, 19\}, \{3, 10\}, \{7, 18\}, \{2, 19\}, \{6, 18\}, \{6, 14\}, \{9, 15\}, \{8, 17\}, \{3, 13\}, \{7, 12\}, \{12, 13\}, \{12, 14\}, \{3, 8\}, \{2, 13\}, \{5, 17\}, \{15, 17\}, \{1, 16\}, \{9, 10\}]
\\
\hline\hline
(17,23)&40&0.26&&[\{6, 11\}, \{5, 12\}, \{4, 14\}, \{13, 16\}, \{5, 8\}, \{4, 7\}, \{15, 16\}, \{13, 17\}, \{5, 17\}, \{9, 14\}, \{2, 10\}, \{7, 15\}, \{11, 12\}, \{2, 8\}, \{1, 4\}, \{3, 9\}, \{3, 16\}, \{10, 13\}, \{3, 6\}, \{2, 7\}, \{9, 17\}, \{1, 8\}, \{1, 6\}]
\\
\cline{1-3}\cline{5-5}
(20,28)&40&0.29&&[{4, 16\}, \{6, 10\}, \{11, 14\}, \{7, 18\}, \{1, 14\}, \{15, 16\}, \{11, 15\}, \{3, 5\}, \{1, 13\}, \{13, 18\}, \{11, 12\}, \{9, 19\}, \{6, 15}, \{8, 10\}, \{4, 20\}, \{1, 8\}, \{17, 20\}, \{9, 12\}, \{4, 7\}, \{5, 17\}, \{5, 8\}, \{19, 20\}, \{3, 18\}, \{7, 11\}, \{13, 19\}, \{2, 6\}, \{2, 17\}, \{5, 12\}]
\\
\cline{1-3}\cline{5-5}
(22,31)&40&0.29&&[\{1, 2\}, \{12, 19\}, \{11, 21\}, \{9, 14\}, \{20, 22\}, \{14, 15\}, \{9, 20\}, \{8, 13\}, \{21, 22\}, \{5, 12\}, \{5, 14\}, \{8, 18\}, \{6, 15\}, \{14, 21\}, \{1, 7\}, \{10, 11\}, \{13, 20\}, \{3, 10\}, \{16, 19\}, \{17, 19\}, \{1, 21\}, \{4, 8\}, \{4, 17\}, \{2, 3\}, \{16, 18\}, \{8, 10\}, \{2, 19\}, \{3, 6\}, \{15, 17\}, \{16, 22\}, \{4, 7\}]
\\
\cline{1-3}\cline{5-5}
(23,33)&40&0.3&0.2&[\{12, 13\}, \{12, 15\}, \{11, 21\}, \{4, 22\}, \{6, 10\}, \{3, 4\}, \{21, 23\}, \{8, 14\}, \{8, 20\}, \{5, 15\}, \{2, 7\}, \{19, 21\}, \{16, 18\}, \{6, 17\}, \{1, 15\}, \{2, 14\}, \{5, 10\}, \{6, 18\}, \{13, 19\}, \{7, 18\}, \{13, 14\}, \{4, 17\}, \{9, 12\}, \{16, 22\}, \{7, 23\}, \{16, 20\}, \{20, 21\}, \{1, 11\}, \{2, 3\}, \{10, 11\}, \{17, 19\}, \{3, 5\}, \{9, 17\}]
\\
\cline{1-3}\cline{5-5}
(40,62)&40&0.35&\cite{cylinder}&[\{26, 29\}, \{26, 36\}, \{27, 29\}, \{8, 28\}, \{39, 40\}, \{4, 17\}, \{19, 29\}, \{15, 24\}, \{12, 23\}, \{25, 32\}, \{8, 17\}, \{26, 28\}, \{17, 23\}, \{21, 30\}, \{4, 23\}, \{2, 25\}, \{16, 18\}, \{11, 16\}, \{11, 15\}, \{27, 33\}, \{32, 35\}, \{3, 20\}, \{1, 6\}, \{2, 38\}, \{20, 37\}, \{3, 5\}, \{35, 37\}, \{24, 32\}, \{9, 31\}, \{9, 30\}, \{30, 34\}, \{3, 6\}, \{5, 38\}, \{19, 28\}, \{33, 37\}, \{5, 22\}, \{24, 26\}, \{22, 32\}, \{13, 15\}, \{18, 36\}, \{14, 24\}, \{16, 33\}, \{9, 13\}, \{2, 12\}, \{7, 18\}, \{10, 14\}, \{37, 39\}, \{5, 13\}, \{1, 27\}, \{28, 31\}, \{1, 34\}, \{20, 31\}, \{9, 10\}, \{7, 34\}, \{6, 11\}, \{36, 38\}, \{18, 35\}, \{30, 39\}, \{12, 14\}, \{25, 34\}, \{19, 40\}, \{23, 33\}]
\\
\hline\hline
(26,32)&48&0.19&&[\{21, 22\}, \{7, 11\}, \{19, 21\}, \{15, 24\}, \{12, 25\}, \{11, 15\}, \{25, 26\}, \{23, 24\}, \{11, 15\}, \{8, 25\}, \{6, 26\}, \{2, 23\}, \{9, 18\}, \{4, 23\}, \{4, 10\}, \{4, 8\}, \{10, 18\}, \{3, 8\}, \{14, 21\}, \{5, 21\}, \{22, 26\}, \{2, 13\}, \{16, 19\}, \{9, 14\}, \{1, 5\}, \{13, 14\}, \{10, 12\}, \{1, 20\}, \{17, 19\}, \{20, 26\}, \{7, 17\}, \{16, 18\}]
\\
\cline{1-3}\cline{5-5}
(27,35)&48&0.23&0.17&[\{1, 20\}, \{21, 23\}, \{14, 18\}, \{14, 22\}, \{9, 10\}, \{6, 20\}, \{12, 25\}, \{16, 20\}, \{24, 26\}, \{13, 25\}, \{5, 24\}, \{11, 18\}, \{4, 8\}, \{4, 22\}, \{12, 21\}, \{16, 22\}, \{15, 24\}, \{10, 19\}, \{4, 27\}, \{11, 25\}, \{19, 23\}, \{2, 23\}, \{17, 21\}, \{8, 12\}, \{1, 15\}, \{3, 8\}, \{2, 26\}, \{26, 27\}, \{15, 18\}, \{9, 13\}, \{6, 13\}, \{7, 14\}, \{3, 5\}, \{16, 17\}, \{7, 10\}]
\\
\cline{1-3}\cline{5-5}
(28,37)&48&0.24&\cite{cylinder}&[\{1, 2\}, \{18, 22\}, \{25, 28\}, \{16, 24\}, \{21, 28\}, \{6, 8\}, \{10, 16\}, \{6, 11\}, \{13, 21\}, \{1, 18\}, \{16, 19\}, \{7, 23\}, \{1, 27\}, \{23, 26\}, \{6, 24\}, \{3, 4\}, \{3, 9\}, \{3, 7\}, \{5, 10\}, \{20, 24\}, \{17, 26\}, \{9, 19\}, \{2, 28\}, \{11, 13\}, \{8, 22\}, \{19, 26\}, \{2, 4\}, \{7, 12\}, \{8, 12\}, \{2, 14\}, \{14, 20\}, \{15, 23\}, \{11, 17\}, \{22, 25\}, \{16, 27\}, \{15, 18\}, \{5, 21\}]\\
\cline{1-3}\cline{5-5}
(29,39)&48&0.26&&[\{1, 2\}, \{6, 11\}, \{22, 23\}, \{18, 21\}, \{2, 27\}, \{26, 28\}, \{23, 28\}, \{13, 20\}, \{23, 24\}, \{1, 14\}, \{6, 12\}, \{1, 4\}, \{3, 15\}, \{1, 10\}, \{13, 27\}, \{5, 29\}, \{18, 19\}, \{5, 7\}, \{3, 12\}, \{7, 19\}, \{11, 16\}, \{24, 25\}, \{17, 28\}, \{9, 22\}, \{12, 29\}, \{26, 29\}, \{18, 20\}, \{3, 9\}, \{6, 20\}, \{2, 7\}, \{10, 15\}, \{14, 22\},  \{11, 25\},\{16, 17\}, \{24, 27\}, \{4, 16\}, \{8, 19\}, \{15, 21\}, \{17, 19\}]
\\
\hline
\end{array}\]
\label{tab-struc2-2}
\end{table}


\begin{figure}
\begin{center}
\includegraphics[scale=0.6]{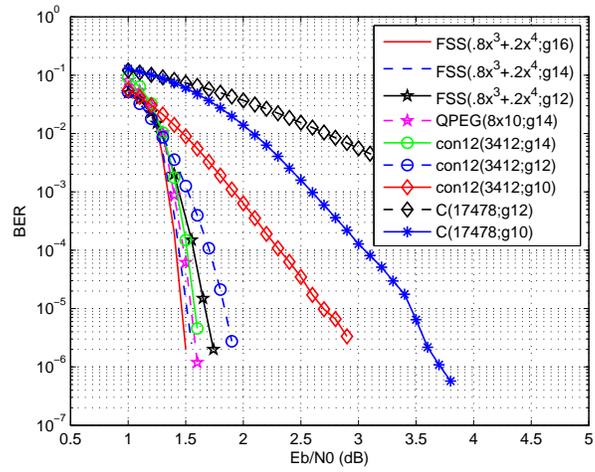}
\end{center}
\caption{\small A comparison between some non-regular FSS codes, QPEG, ${\rm Aff}^*$(16)-based and cylinder-type QC LDPC codes of length 58000 and rate about 0.2.} \label{c3,4}
\end{figure}

\end{document}